\documentclass[twocolumn,prx,aps,amsmath,amssymb,longbibliography]{revtex4-1}

\usepackage{graphicx}
\usepackage{dcolumn}
\usepackage{bm}

\begin{document}

\title{Majorana entropy revival via tunneling phases}

\author{Sergey Smirnov}
\affiliation{P. N. Lebedev Physical Institute of the Russian Academy of
  Sciences, 119991 Moscow, Russia}
\email{1) sergej.physik@gmail.com\\2)
  sergey.smirnov@physik.uni-regensburg.de\\3) ssmirnov@sci.lebedev.ru}

\date{\today}

\begin{abstract}
Measuring the Majorana entropy $S_M=k_B\log(2^\frac{1}{2})$ may uniquely
reveal whether an initial equilibrium state of a nanoscale device is of
Majorana nature and subsequent operations deal with an essentially nonlocal
pair of non-Abelian Majorana bound states and not with trivial or other
accidental non-Abelian states. However, in realistic setups both Majorana
modes are inevitably involved in tunneling processes. We show that even when
the tunneling amplitude of one Majorana mode is significantly suppressed, the
Majorana entropy ruins and straightforward experiments will in general detect
entropy $S\ll S_M$. To avoid this general problem we present a mechanism of
the Majorana entropy revival via the tunneling phases of the Majorana modes
and demonstrate that to successfully observe the universal Majorana plateau
$S=S_M$ one should intelligently tune the tunneling phases instead of leaving
them uncontrolled. Practical feasibility of appropriate Majorana entropy
measurements is supported by an example with parameters well achievable in
modern labs.
\end{abstract}

\maketitle
\section{Introduction}\label{Intro}
Quantum thermodynamic properties of a nanoscopic system offer a rich source of
information about its physical states. For instance, the system's entropy at
low temperatures provides the physical nature of the quantum ground states
which also determine quantum transport characteristics such as the linear
conductance. Thus getting access to thermodynamics of a nanoscopic system is
an extremely challenging task for modern experiments to reveal both the
system's quantum ground states and how they govern quantum transport response
of the system.

This is particularly important for nanoscopic systems involving
one-dimensional topological superconductors in their topologically nontrivial
phases replicating
\cite{Alicea_2012,Flensberg_2012,Sato_2016,Aguado_2017,Lutchyn_2018} the
prototypical phase of the Kitaev model \cite{Kitaev_2001} realized, {\it e.g},
on the basis of semiconductors with strong spin-orbit interactions
\cite{Lutchyn_2010,Oreg_2010}. This phase is characterized by non-Abelian
Majorana bound states (MBSs) localized at the ends of a topological
superconductor appearing after a topological quantum phase transition as a
consequence of a qualitative change of the quantum ground state from a trivial
one to a nontrivial Majorana state.

The Majorana quantum ground state dictates specific features of quantum
transport response such as the linear conductance observed in experiments
\cite{Mourik_2012,Albrecht_2016}, in particular, its universality 
\cite{Zhang_2018}. In fact, quantum transport predicts many specific Majorana
induced signatures in various characteristics such as thermoelectric currents
\cite{Leijnse_2014,Ramos-Andrade_2016,Hong_2020,Chi_2020,Bondyopadhaya_2020},
shot noise
\cite{Liu_2015,Liu_2015a,Haim_2015,Valentini_2016,Smirnov_2017,Bathellier_2019},
quantum noise \cite{Smirnov_2019}, thermoelectric shot noise
\cite{Smirnov_2018}, thermoelectric quantum noise \cite{Smirnov_2019a}, tunnel
magnetoresistances in ferromagnetic systems \cite{Tang_2020}, linear
conductances in quantum dissipative systems \cite{Zhang_2020}, quantum
transmission in photon-assisted transport \cite{Chi_2020a}, pumped heat and
charge statistics \cite{Simons_2020}. Transport properties may also be
combined with thermodynamic ones \cite{Smirnov_2020a} to observe dual Majorana
universality.

Revealing many consequences of the Majorana quantum ground states, quantum
transport is, however, unable to directly access the Majorana ground states
themselves. To uniquely detect MBSs within quantum transport experiments one
should be able to adapt them to mimic Majorana braiding protocols, {\it e.g.},
by means of nonequilibrium noise measurements \cite{Manousakis_2020}.

Nevertheless, it is highly appealing to uniquely reveal MBSs directly from the
system's quantum ground state avoiding braiding protocols or their conceivable
counterparts. Successful experimental entropy measurements
\cite{Hartman_2018,Kleeorin_2019} provide an exceptional opportunity to access
the Majorana entropy in nanoscopic systems. In practice, however, this is not
a straightforward task. Indeed, to measure the Majorana entropy one could
consider an ideal setup presumably involving a pair of highly nonlocal
non-Abelian MBSs with a finite coupling to only one Majorana mode ignoring
completely any coupling to the second Majorana mode as proposed in
Refs. \cite{Smirnov_2015,Sela_2019}.

Here we demonstrate that the above idealization will likely fail to guide
experimental observations of the Majorana entropy
$S_M=k_B\log(2^\frac{1}{2})$ because of inevitable finite coupling to the
second Majorana mode in a realistic setup. As discussed below, even if the
tunneling amplitude of the second Majorana mode is several orders of magnitude
smaller than the tunneling amplitude of the first Majorana mode,
straightforward experiments will be essentially brought in a regime with
entropy $S\ll S_M$. We reveal that such a strong sensitivity of the system's
entropy to the coupling of the second Majorana mode is the result of emergence
of additional degrees of freedom, namely the Majorana tunneling
phases. Exactly these new degrees of freedom, if left uncontrolled, ruin the
Majorana entropy to a much smaller value. Remarkably, it turns out that
exactly the Majorana tunneling phases provide a revival of the Majorana
entropy that is a return of the system to the Majorana quantum ground state
with the entropy $S=S_M$. As we show, one may revive Majorana quantum ground
states via tunneling phases or a gate voltage in setups with experimentally
realizable parameters.
\begin{figure}
\includegraphics[width=8.0 cm]{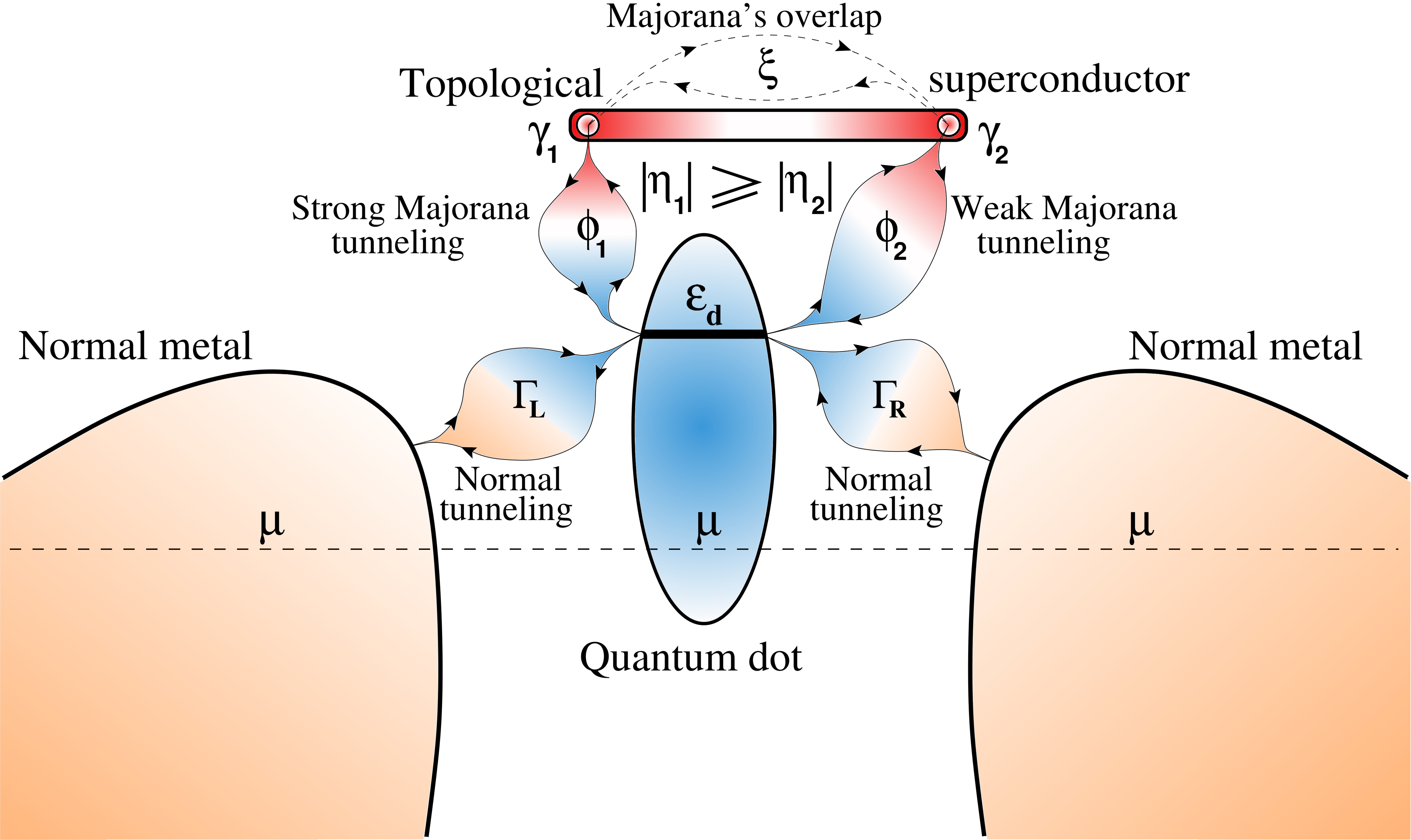}
\caption{\label{figure_1} A nanoscopic system whose quantum thermodynamic and
  transport properties stem from the nonlocal nature of non-Abelian MBSs.}
\end{figure}

The paper is organized as follows. In Section \ref{TMME} we present a
realistic nanoscopic setup where both Majorana modes are involved in tunneling
processes with corresponding tunneling phases which become new parameters of
the system's entropy. The dependence of the system's entropy on the tunneling
phases is numerically explored in Section \ref{Res} which shows how the
Majorana ground state revives via tuning the tunneling phases restoring
simultaneously the Majorana values of the entropy and linear
conductance. Additionally, we demonstrate how to revive the Majorana state via
a gate voltage at fixed arbitrary values of Majorana tunneling phases. We
conclude with Section \ref{Concl} where we estimate a possibility of an
experimental implementation of the Majorana ground state revival, mention
connection to driven dissipative protocols using tunneling phases to control
Majorana qubits and also provide an outlook on various systems where Majorana
tunneling phases might play an essential role.
\section{Theoretical model and Majorana entropy}\label{TMME}
To specify our discussion let us consider a concrete setup shown in
Fig. \ref{figure_1}. The system is composed of a quantum dot, two normal
massive metals and a grounded topological superconductor with MBSs localized
at its ends. The quantum dot interacts with the normal metals and topological
superconductor through tunneling mechanisms.

The quantum dot Hamiltonian is
\begin{equation}
  \hat{H}_d=\epsilon_dd^\dagger d.
  \label{Ham_QD}
\end{equation}
The location of the energy level $\epsilon_d$ with respect to the chemical
potential $\mu$ is tuned by a gate voltage. Both the energy level $\epsilon_d$
and the chemical potential $\mu$ are measured with respect to the middle of
the induced superconducting gap at which the Majoranas are bound. Below we
focus on the case when the chemical potential coincides with the middle of the
induced superconducting gap. Therefore varying $\epsilon_d$ by a gate voltage
with respect to the middle of the induced superconducting gap is identical to
varying it with respect to the chemical potential $\mu$ which has a unique
value in equilibrium where it only makes sense to explore the system's
entropy. Thus, in what follows, values of $\epsilon_d$ will be assumed to be
given with reference to the chemical potential $\mu$. While the impact of
deviations of the chemical potential from the middle of the induced
superconducting gap will be explored in future research, the case we explore
here is also adopted in many other works. For example, in the limit of only
one Majorana mode coupled to the quantum dot our Green's functions given below
are similar to those in Refs. \cite{Liu_2015,Liu_2015a} for the noninteracting
setups. In particular, in this limit our Green's functions result in the same
values of the linear conductance and shot noise (see
Refs. \cite{Smirnov_2015,Smirnov_2017}) as those obtained in
Refs. \cite{Liu_2015,Liu_2015a}.

The Hamiltonian of the normal metals,
\begin{equation}
  \hat{H}_c=\sum_{l=L,R}\sum_k\epsilon_kc^\dagger_{lk}c_{lk},
  \label{Ham_Cont}
\end{equation}
is a sum over the left ($L$) and right ($R$) contacts as well as over the
momentum index $k$. It is characterized by continuous spectra $\epsilon_k$
resulting in a density of states $\nu(\epsilon)$ assumed energy independent in
the vicinity of the chemical potential, $\nu(\epsilon)\approx\nu_c/2$. The
normal metals are assumed to be in equilibrium specified by the Fermi-Dirac
distribution with the chemical potentials $\mu_{L,R}$ and temperature $T$,
$f_{L,R}(\epsilon)=\{\exp[(\epsilon-\mu_{L,R})/k_BT]+1\}^{-1}$. Below the
system's entropy is calculated in equilibrium, $\mu_L=\mu_R=\mu$, whereas we
calculate the linear conductance for $\mu_{L,R}=\mu\pm eV/2$, where $V$ is an
infinitesimal bias voltage between the two normal metals, $V\rightarrow 0$.

Here it is appropriate to mention that while the electric current induced by
the bias voltage $V$ is a nonequilibrium characteristic, the linear
conductance is a transport property fully specified by the system's
equilibrium state. Therefore at low temperatures the behavior of the linear
conductance is governed by the system's quantum ground state whose nature is
revealed by the system's entropy. Below we will demonstrate that the linear
conductance takes the Majorana fractional value when the entropy signals that
the system's quantum ground state has acquired the Majorana nature. 

The tunneling interaction of the quantum dot with the left and right normal
metals is described by the Hamiltonian
\begin{equation}
  \hat{H}_{d-c}=\sum_{l=L,R}\mathcal{T}_l\sum_kc^\dagger_{lk}d+\text{H.c.}
  \label{Ham_tun_QD_cont}
\end{equation}
bringing the energy scale $\Gamma\equiv\Gamma_L+\Gamma_R$,
$\Gamma_l\equiv\pi\nu_C|\mathcal{T}_l|^2$. Here we have assumed that the
tunneling matrix elements do not depend on the momentum index $k$. Physically
this simplification corresponds to a scabrous tunneling barrier where the
tunneling is most intensive at only one point of the barrier, namely at its
thinnest point. Under such physical conditions our simplification is well
justified. In the opposite situation of a smooth tunneling barrier having
approximately constant width our simplification will not affect the results at
low temperatures when the range of relevant energies is very small. Our
simplification would probably become less precise for a smooth tunneling
barrier at high temperatures when the MBSs are no longer effective. However,
at low temperatures, when the Majorana low energy sector governs the system's
physical behavior, our results will not be much influenced by that
simplification even for a smooth tunneling barrier. We also note that in
equilibrium, when $\mu_L=\mu_R=\mu$, the two normal metals are
indistinguishable and are physically equivalent to a single normal metal. As a
result, the system's entropy must not depend on specific values of the
energies $\Gamma_L$ and $\Gamma_R$ but it must depend only on their sum that
is on $\Gamma$. In contrast, the value of the linear conductance may depend on
specific values of $\Gamma_L$ and $\Gamma_R$. Below, when we calculate the
linear conductance, we assume $\Gamma_L=\Gamma_R=\Gamma/2$.

The highly nonlocal Majorana modes localized at the ends of the topological
superconductor, $\gamma_1$ and $\gamma_2$, are both linked with the quantum
dot and may also have a finite overlap between each other. The links of
$\gamma_1$ and $\gamma_2$ with the quantum dot have, respectively, tunneling
amplitudes $|\eta_1|$ and $|\eta_2|$ as well as tunneling phases $\phi_1$ and
$\phi_2$. The corresponding
Hamiltonian is
\begin{equation}
  \hat{H}_{d-ts}=\eta_1^*d^\dagger\gamma_1+\eta_2^*d^\dagger\gamma_2+\text{H.c.},
  \label{Ham_tun_QD_TS}
\end{equation}
where $\gamma_i^\dagger=\gamma_i$, $\{\gamma_i,\gamma_j\}=2\delta_{ij}$,
$\eta_{1,2}=|\eta_{1,2}|\exp(i\phi_{1,2})$. The Majorana mode $\gamma_1$ is
linked to the quantum dot stronger than the Majorana mode $\gamma_2$,
{\it i.e.} $|\eta_1|>|\eta_2|$. The Majorana's overlap, shown by the dashed
arrows, is modeled by the Hamiltonian
\begin{equation}
  \hat{H}_{ts}=i\xi\gamma_2\gamma_1/2
  \label{Ham_TS}
\end{equation}
with the overlap energy scale $\xi$. We note that physical observables, in
particular the system's entropy, cannot depend separately on the two phases
$\phi_1$ and $\phi_2$ but they must depend only on their difference
$\Delta\phi\equiv\phi_1-\phi_2$.

As in Ref. \cite{Smirnov_2015}, replacing the second quantized operators in
the above Hamiltonians with the corresponding Grassmann fields on the
imaginary time axis, the thermodynamic partition function $Z$ of the system is
represented by a field integral in imaginary time with the Grassmann fields
subject to the antiperiodic boundary conditions \cite{Altland_2010}. The
system's entropy is then obtained from the thermodynamic potential
$\Omega=-k_BT\log Z$ as the first derivative over the temperature,
$S=-\partial\Omega/\partial T$. It has the form:
\begin{equation}
  \begin{split}
    &S=k_B\log\biggl[\cosh\biggl(\frac{\xi}{2k_BT}\biggl)\biggl]-\frac{\xi}{2T}\tanh\biggl(\frac{\xi}{2k_BT}\biggl)\\
    &+k_B\log(2)+\frac{1}{8\pi k_BT^2}\int_{-\infty}^\infty d\epsilon\frac{\epsilon\phi(\epsilon)}{\cosh^2(\frac{\epsilon}{2k_BT})},
  \end{split}
  \label{Entropy}
\end{equation}
where $\phi(\epsilon)$ represents the phase of a complex expression involving
the retarded and advanced hole-particle, hole-hole and particle-particle
Green's functions,
\begin{equation}
  G_{hp}^A(-\epsilon)G_{hp}^R(\epsilon)-G_{hh}^A(-\epsilon)G_{pp}^R(\epsilon)=\rho(\epsilon)\exp[i\phi(\epsilon)],
  \label{Expr_GF}
\end{equation}
where $iG_{jj'}^{R,A}(t|t')\equiv\pm\Theta(\pm t\mp t')\langle\{d_j(t),d_{j'}(t')\}\rangle$,
$j=p,h$ and $d_p\equiv d^\dagger$, $d_h\equiv d$. The last term in
Eq. (\ref{Entropy}) takes into account the tunneling interactions of the
quantum dot with the normal metals and topological superconductor while the
first three terms may be interpreted as the entropy of a two-level system in
which the distance between the two energy levels is equal to $\xi$.

The Green's functions depend on the parameters of the above setup, in
particular on the tunneling phase difference $\Delta\phi$, and are found from
a field integral in real time, the Keldysh field integral
\cite{Altland_2010}. The retarded Green's functions have the following form:
\begin{equation}
  G_{ij}^R(\epsilon)=\frac{g_{ij}^R(\epsilon)}{g^R(\epsilon)},
  \label{R_GF}
\end{equation}
where
\begin{equation}
  \begin{split}
    &g_{hp}^R(\epsilon)=2\hbar\bigl(\epsilon^2-\xi^2\bigr)\bigl[i\Gamma+2(\epsilon_d+\epsilon)\bigr]\\
    &-8\hbar\bigl[2\xi|\eta_1||\eta_2|\sin(\Delta\phi)+\epsilon(|\eta_1|^2+|\eta_2|^2)\bigr],\\
    &g_{pp}^R(\epsilon)=-8\hbar\epsilon(\eta_1^2+\eta_2^2),\\
    &g_{hh}^R(\epsilon)=-8\hbar\epsilon\bigl[(\eta_1^*)^2+(\eta_2^*)^2\bigr],\\
    &g^R(\epsilon)=(\xi^2-\epsilon^2)\bigl[4\epsilon_d^2+(\Gamma-2i\epsilon)^2\bigr]\\
    &+64|\eta_1|^2|\eta_2|^2\sin^2(\Delta\phi)+32\epsilon_d\xi|\eta_1||\eta_2|\sin(\Delta\phi)\\
    &-8i\epsilon(\Gamma-2i\epsilon)(|\eta_1|^2+|\eta_2|^2).
  \end{split}
  \label{R_GF_hp}
\end{equation}
The advanced Green's functions are obtained from the relations:
\begin{equation}
  \begin{split}
    &G_{hp}^A(\epsilon)=\bigl[G_{hp}^R(\epsilon)\bigr]^*,\\
    &G_{pp}^A(\epsilon)=\bigl[G_{hh}^R(\epsilon)\bigr]^*,\\
    &G_{hh}^A(\epsilon)=\bigl[G_{pp}^R(\epsilon)\bigr]^*,
  \end{split}
  \label{A_GF}
\end{equation}
which follow from the definitions of the retarded and advanced hole-particle,
hole-hole and particle-particle Green's functions given above.
\section{Results}\label{Res}
\begin{figure}
\includegraphics[width=8.0 cm]{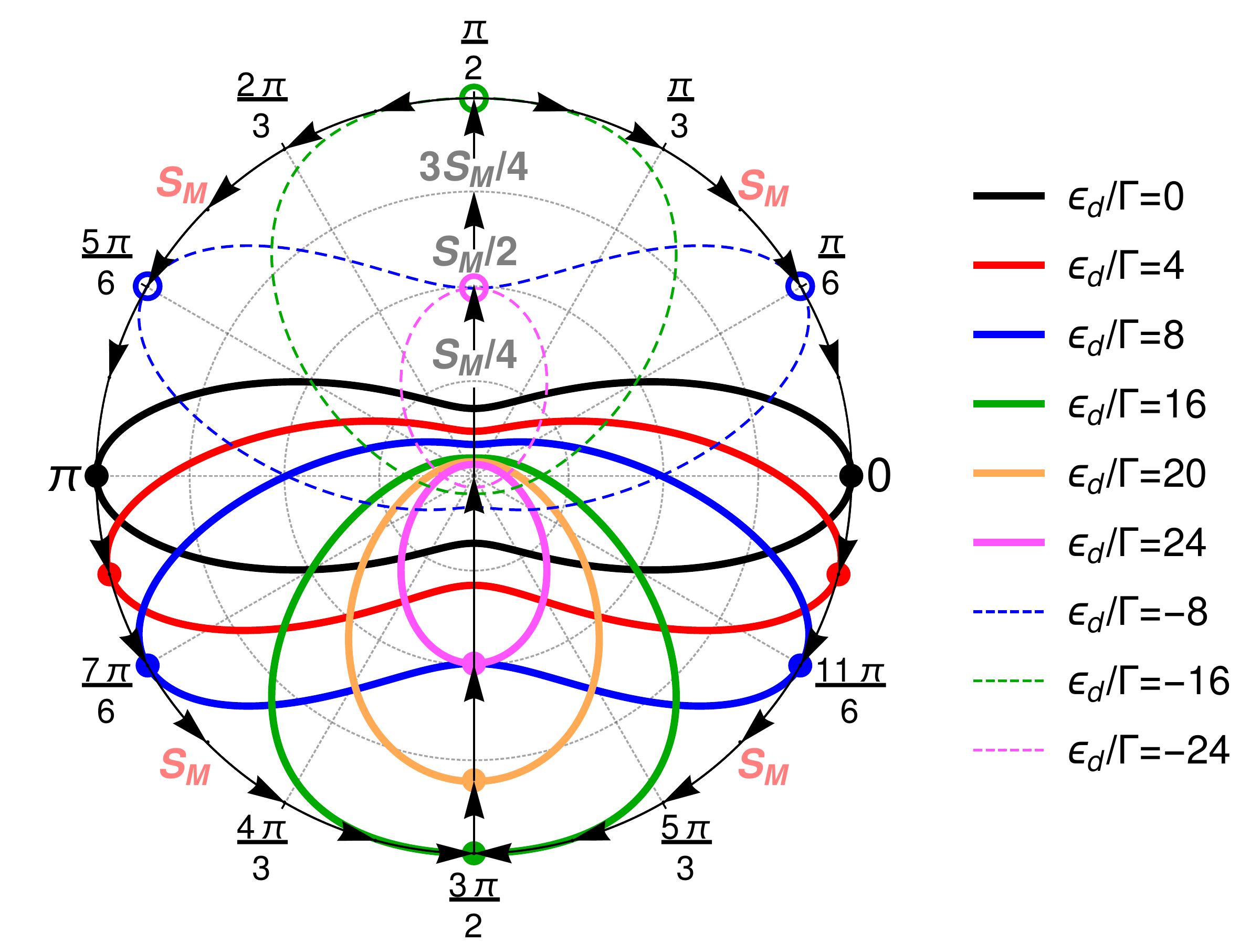}
\caption{\label{figure_2} A plot representing the system's entropy $S$
  as a function of the difference of the tunneling phases,
  $\Delta\phi=\phi_1-\phi_2$, in polar coordinates. Here the splitting energy
  in Eq. (\ref{Eps_M}) is $\epsilon_M/\Gamma=16$ and thus a single entropy
  maximum splits into two maxima at $\epsilon_d=\pm\epsilon_M$, that is at
  $\epsilon_d/\Gamma=\pm 16$ (solid and dashed green curves, respectively).}
\end{figure}

In Fig. \ref{figure_2} we show the results obtained for the system's entropy
as a function of the tunneling phase difference $\Delta\phi$ in polar
coordinates. Specifically, the distance from the center (the origin of
coordinates) to a point on a curve is equal to $S$ while the polar angle is
equal to $\Delta\phi$. Different curves correspond to different values of a
gate voltage controlling $\epsilon_d$. The solid red, blue, green, orange and
magenta curves show $S$ for positive values of $\epsilon_d$ while the dashed
blue, green and magenta curves show $S$ for the corresponding negative values
of $\epsilon_d$. The solid black curve is for $\epsilon_d=0$. Here
$k_BT/\Gamma=10^{-8}$, $|\eta_1|/\Gamma=4\cdot 10^2$,
$|\eta_2|/\Gamma=10^{-4}$, $\xi/\Gamma=10^{-2}$. As can be seen, although
$|\eta_1|$ is more than six orders of magnitude larger than $|\eta_2|$, all
the curves are highly anisotropic showing a very strong dependence of the
system's entropy $S$ on the tunneling phase difference $\Delta\phi$. The
points on the curves with $\epsilon_d\geqslant 0$ and the circles on the
curves with $\epsilon_d<0$ indicate where $S$ reaches its maximal value. The
radius of the largest polar circle is equal to the Majorana entropy
$S_M$.

We find numerically that all the curves with $|\epsilon_d|<\epsilon_M$
touch the largest polar circle $S_M$ at two points while the curves with
$|\epsilon_d|=\epsilon_M$ touch it only at one point, at $\Delta\phi=\pi/2$
for $\epsilon_d<0$ and at $\Delta\phi=3\pi/2$ for $\epsilon_d>0$.

Here there appears a new energy scale,
\begin{equation}
  \epsilon_M\equiv\frac{4|\eta_1||\eta_2|}{\xi},
  \label{Eps_M}
\end{equation}
which has been identified numerically for setups with
$|\eta_2|\ll|\eta_1|$. It is interesting to note that this energy scale and
the above mentioned entropy maximum points do not appear in our numerical
calculations for setups with $|\eta_2|\sim|\eta_1|$ which are not explored
here but will be a topic of our future research.

For $|\epsilon_d|>\epsilon_M$ the maximal values of $S$ start to depend on
$\epsilon_d$ but do not rotate any more and remain at the two fixed polar
angles, $\Delta\phi=\pi/2$ ($\epsilon_d<0$) and $\Delta\phi=3\pi/2$
($\epsilon_d>0$). When $|\epsilon_d|\rightarrow\infty$, the maximal values of
$S$ at these two polar angles decrease and go from the universal Majorana
value $S_M$ to zero. This situation is demonstrated by the black arrows
representing a flow of the entropy maximum in the direction of increasing
values of $\epsilon_d$. For large negative values of $\epsilon_d$ one starts
at the center and moves up along the $y$-axis when $\epsilon_d$ increases up
to $\epsilon_d=-\epsilon_M$ where the maximal value of $S$ is equal to the
Majorana entropy $S_M$. After this point, when $\epsilon_d$ increases further,
the maximal value of $S$ does not depend on $\epsilon_d$ and remains equal to
the universal Majorana entropy $S_M$. However, the polar angle at which it is
observed splits from $\Delta\phi=\pi/2$ into two polar angles which rotate,
upon increasing $\epsilon_d$, on the largest polar circle $S_M$ in the
opposite directions, anticlockwise and clockwise, and merge again at the polar
angle $\Delta\phi=3\pi/2$ when $\epsilon_d=\epsilon_M$. For
$\epsilon_d>\epsilon_M$ the maximal value of $S$ again starts to depend on
$\epsilon_d$ and goes from the universal Majorana value $S_M$ to zero when
$\epsilon_d$ goes to large positive values, that is one moves up along the
$y$-axis from the largest polar circle $S_M$ to the center where the flow
returns to its starting point and finally stops.
\begin{figure}
\includegraphics[width=8.0 cm]{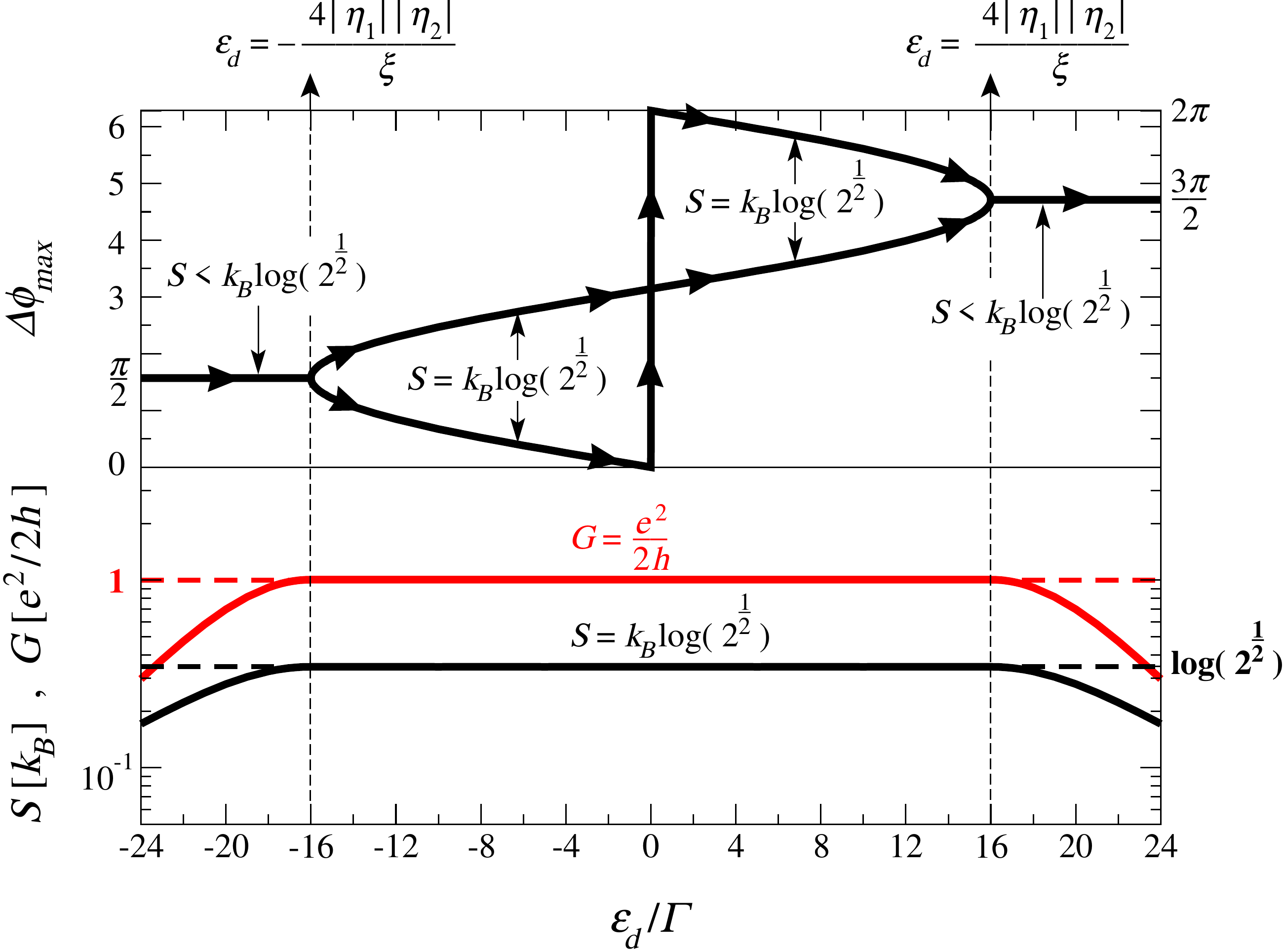}
\caption{\label{figure_3} Probing Majorana universality in quantum
  thermodynamic and transport behavior governed by nonlocal MBSs.}
\end{figure}

In Fig. \ref{figure_3} we demonstrate the universality of the Majorana ground
state. The parameters have the same values as in Fig. \ref{figure_2}. Upper
panel: The differences of the tunneling phases $\Delta\phi$ corresponding to
maximal values of the system's entropy $S$ are shown as functions of
$\epsilon_d$. As in Fig. \ref{figure_2}, the black arrows indicate the flow of
the maximal value of $S$ in the direction of increasing values of
$\epsilon_d$. The flow splits at the point $\epsilon_d=-\epsilon_M$ into two
flows which merge again at the point $\epsilon_d=\epsilon_M$. For $\Delta\phi$
on the two flows inside the window $|\epsilon_d|\leqslant\epsilon_M$ the
system's entropy is equal to the Majorana value, $S=S_M$, and it does not
depend on $\epsilon_d$ revealing universal thermodynamic behavior induced by
MBSs. Outside that universal window, that is for $|\epsilon_d|>\epsilon_M$,
the system's entropy on the flow is no longer universal, that is it depends on
$\epsilon_d$ and its value always remains below the Majorana value,
$S<S_M$. Lower panel: The black curve is the entropy maximum reached at
$\Delta\phi$ from the upper panel. In other words, this curve shows the
system's entropy on the flow shown at the upper panel as a function of
$\epsilon_d$. It has the Majorana plateau $S=S_M$ for
$|\epsilon_d|\leqslant\epsilon_M$ and decreases when moving away from this
universal window. The red curve shows the behavior of the maximal value of the
linear conductance. The maximum of the linear conductance is reached also at
$\Delta\phi$ from the upper panel. Therefore, the linear conductance is also
taken on the flow from the upper panel. The linear conductance has the
Majorana plateau $G=G_M\equiv e^2/2h$ in the same range of $\epsilon_d$ where
$S=S_M$. This clearly shows how the nontrivial Majorana ground state, which
stores inside the nonlocality of the MBSs, uniquely determines the quantum dot
linear response.
\begin{figure}
\includegraphics[width=8.0 cm]{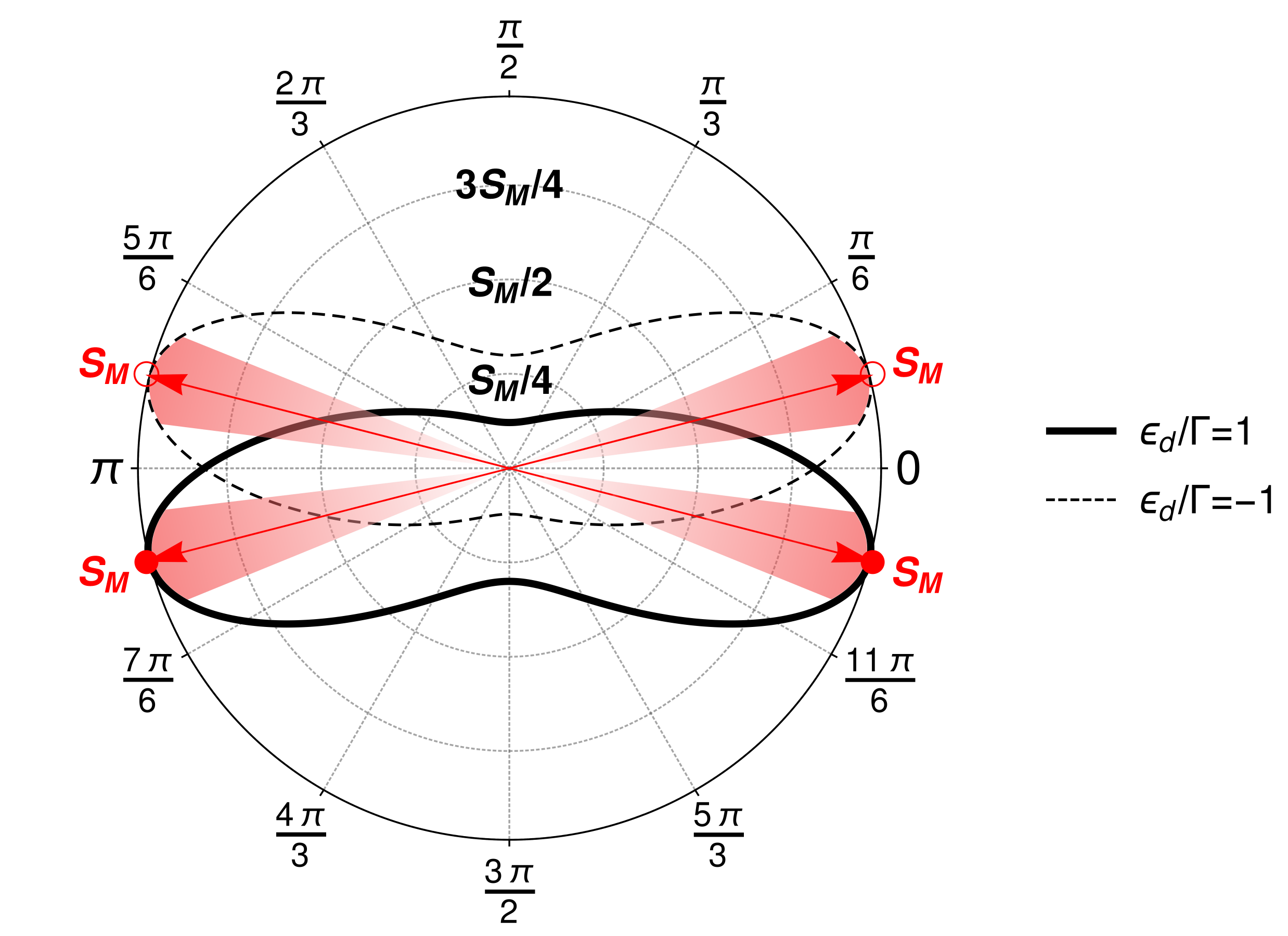}
\caption{\label{figure_4} A polar plot demonstrating persistence of the
  Majorana anisotropic character of the system's entropy at high
  temperatures.}
\end{figure}
\begin{figure}
\includegraphics[width=8.0 cm]{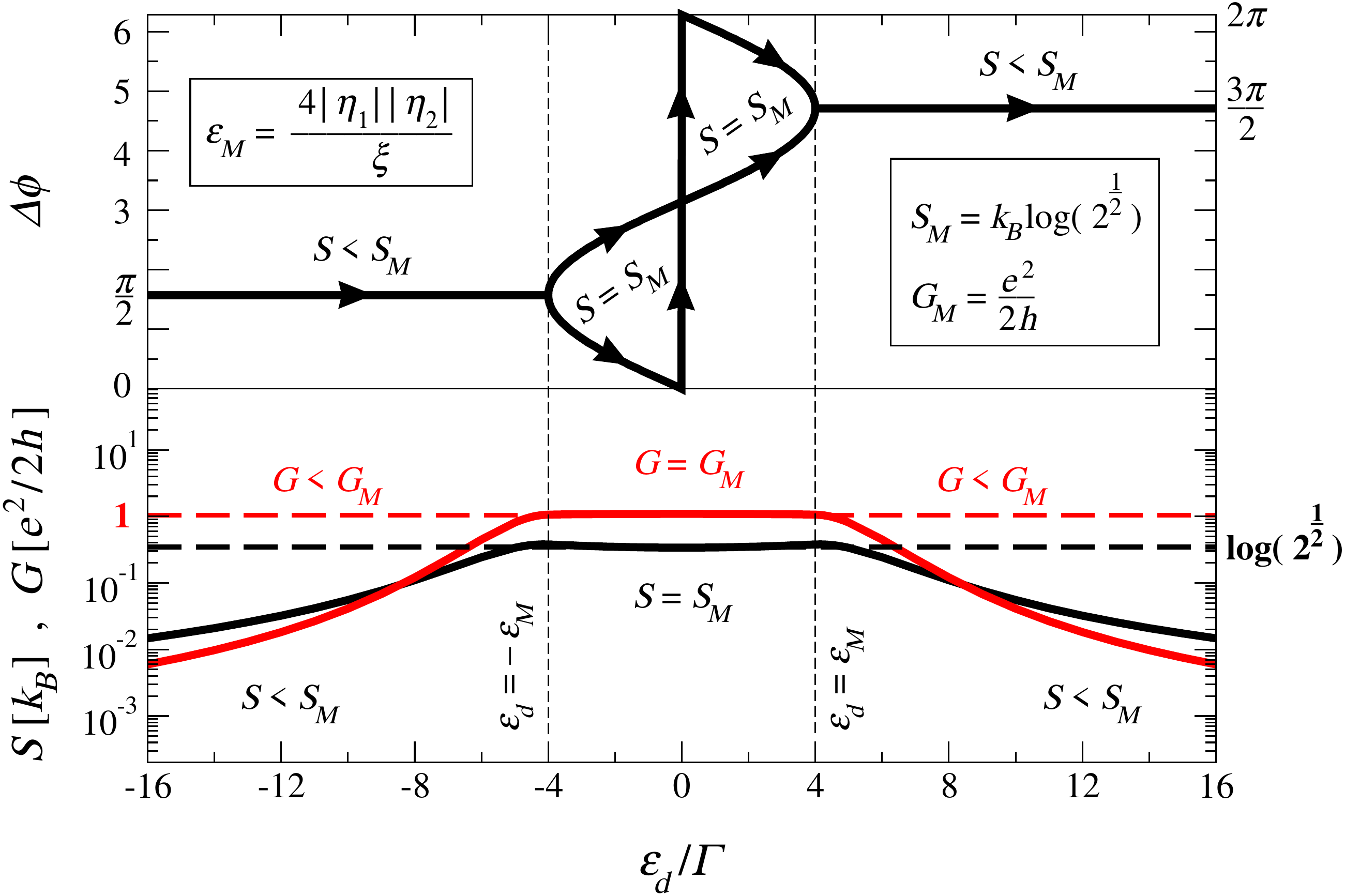}
\caption{\label{figure_5} Persistence of the Majorana universality of the
  system's entropy $S$ and linear conductance $G$ at high temperatures.}
\end{figure}

As shown in Fig. \ref{figure_4}, a strong anisotropy of the system's entropy
is observed also after an essential increase of the temperature at
experimentally relevant values of the parameters. Here $k_BT/\Gamma=10^{-2}$,
$|\eta_1|/\Gamma=1$, $|\eta_2|/\Gamma=10^{-1}$, $\xi/\Gamma=10^{-1}$. The
solid curve is for $\epsilon_d/\Gamma=1$ and the dashed curve is for
$\epsilon_d/\Gamma=-1$. The red arrows correspond to $\Delta\phi$ where
$S=S_M$. The red angular sectors display vicinities of $\Delta\phi$ where $S$
is close to $S_M$. As can be seen, the anisotropic character and the maximal
value $S_M$ of the system's entropy are revealed even when the temperature has
been raised up six orders of magnitude in comparison with
Fig. \ref{figure_2}.
\begin{figure}
\includegraphics[width=8.0 cm]{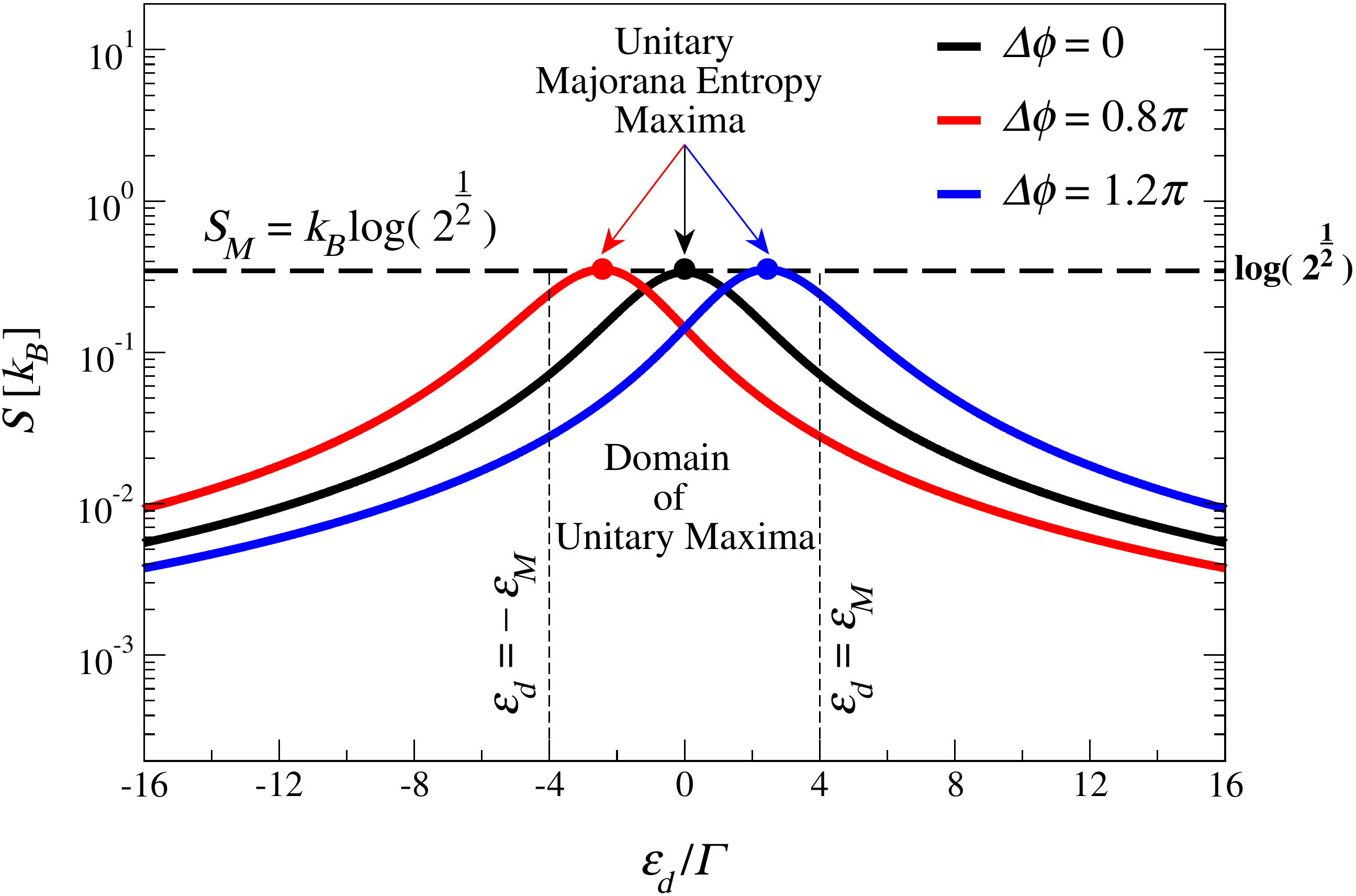}
\caption{\label{figure_6} Second type of experiment: revival and detection of
  the Majorana entropy via varying the quantum dot gate voltage $\epsilon_d$
  and at the same time keeping the tunneling phase difference $\Delta\phi$
  fixed.}
\end{figure}

Universal Majorana behavior of $S$ and $G$ is also observed at high
temperatures as demonstrated in Fig. \ref{figure_5}. All the parameters have
the same values as in Fig. \ref{figure_4}. Upper panel: The flow of
$\Delta\phi$ on which $S$ and $G$ reach their maximal values. The flow is
parameterized by $\epsilon_d$. Lower panel: The black and red curves show,
respectively, $S$ and $G$ on the flow from the upper panel. As can be seen,
the universal Majorana thermodynamic and transport behavior is robust and
retains all its specific features even at very high temperatures. In
particular, the universal Majorana plateaus $S=S_M$ and $G=G_M$ are
simultaneously developed inside the window
$|\epsilon_d|\leqslant\epsilon_M$. Here, for the given set of parameters, this
universal window is narrowed four times in comparison with
Fig. \ref{figure_3}.

We note that, as can be seen from Figs. \ref{figure_2} and \ref{figure_4}, for
a given gate voltage the Majorana entropy is ruined, $S\ll S_M$, everywhere
except for two values of the phase difference $\Delta\phi_{max}$, where
$S=S_M$ and two narrow angular sectors around $\Delta\phi_{max}$ where
$S\approx S_M$. Such behavior provides two types of experimental detection of
the Majorana entropy.

First, it is clear that experiments which keep the gate voltage fixed and at
the same time do not control the Majorana tunneling phases will reveal with
high probability the system's entropy $S\ll S_M$. However, in experiments
tuning the Majorana tunneling phases one will detect a maximum, $S_{max}$, of
the system's entropy as a function of $\Delta\phi$. Observing $S_{max}=S_M$
will be a fully conclusive signature of the topologically nontrivial Majorana
quantum ground state which has been revived via the corresponding tunneling
phases.

Second, in a setup where the Majorana tunneling phases are fixed because, due
to some reasons, it is difficult to vary their values one has to keep in mind
that the tunneling phase difference with an accidental finite value is in
general induced during the preparation of this experimental setup. Under such
circumstances we suggest to vary the quantum dot gate voltage. This is a well
established experimental technique often used in more traditional
experiments. Results presented in Figs. \ref{figure_2} and \ref{figure_4}
suggest that for a fixed value of $\Delta\phi$ in experiments varying the gate
voltage one will detect a maximum, $S_{max}$, of the system's entropy as a
function of $\epsilon_d$. This is demonstrated in Fig. \ref{figure_6} where
the three solid curves correspond to three different values of the tunneling
phase difference: $\Delta\phi=0$ (black), $\Delta\phi=0.8\pi$ (red) and
$\Delta\phi=1.2\pi$ (blue). The other parameters have the same values as in
Fig. \ref{figure_4}. In this type of experiment the system's entropy will
reach its maximum $S_{max}$ at some value of the gate voltage
$\epsilon_d=\epsilon_{d,max}$ located in the domain
$-\epsilon_M\leqslant\epsilon_{d,max}\leqslant\epsilon_M$ as shown in
Fig. \ref{figure_6}. Detecting $S_{max}=S_M$ will be a totally conclusive
proof of the topologically nontrivial Majorana quantum ground state which in
this type of experiment has been revived via the quantum dot gate
voltage. Additionally, in this type of experiment observing
$\epsilon_{d,max}\neq 0$ indicates that the tunneling phase difference is
finite, $\Delta\phi\neq 0$. This information is important to properly analyze
possible subsequent transport experiments in the same setup because they may
strongly depend on $\Delta\phi$ as it happens, for example, with the linear
conductance discussed above.
\section{Conclusion}\label{Concl}
In conclusion, let us estimate the experimental relevance of the parameters
used for Figs. \ref{figure_4}-\ref{figure_6}. Obviously, $|\epsilon_d|$ should
not exceed the induced superconducting gap $\Delta$. For the largest energy
scale $\Gamma$ we take $\Gamma\approx\Delta$. Thus one should expect to
observe the Majorana universality in the window
$|\epsilon_d|\leqslant\Gamma$. For the parameters $|\eta_1|$, $|\eta_2|$ and
$\xi$ used to obtain the results shown in Figs. \ref{figure_4}-\ref{figure_6}
one has $\epsilon_M=4\Gamma$ and thus the universal window is
$|\epsilon_d|\leqslant 4\Gamma$. However, if necessary, one can properly
reduce the width of the universal window changing the above parameters. For
example, we have numerically checked that decreasing the parameter $|\eta_2|$
four times, $|\eta_2|/\Gamma=2.5\cdot 10^{-2}$ ($\epsilon_M=\Gamma$, according
to Eq. (\ref{Eps_M})), leads to results similar to those shown in
Figs. \ref{figure_4}-\ref{figure_6} but with the universal window narrowed
four times, $|\epsilon_d|\leqslant\Gamma$. Concerning the temperature, for
$\Delta\approx 250\,\mu\text{eV}$ (see Ref. \cite{Mourik_2012}) one obtains
$T\approx 30\,\text{mK}$ as the upper limit. This means that in experiments
performed at $T\gtrapprox 30\,\text{mK}$ quasiparticles above the induced
superconducting gap will start to contribute to the system's macroscopic
states leading to an increase of the entropy. However, for temperatures
$T\lessapprox 30\,\text{mK}$ the contribution from quasiparticles above the
induced superconducting gap will be significantly suppressed and thus the
entropy will be essentially governed by MBSs resulting in the behavior
revealed in this work. This can be seen from the expression for the entropy,
Eq. (\ref{Entropy}), where the hyperbolic cosine exponentially suppresses
contributions from energies $\epsilon$ with $|\epsilon|\gtrsim k_BT$. Since in
our work we always have $k_BT\ll\Gamma$, contributions from quasiparticles
above the induced superconducting gap are exponentially suppressed (with a
rough estimate given by $\exp(-\Gamma/2k_BT)$). The latter argument is also
justified by the fact that in our noninteracting case the retarded and
advanced Green's functions, Eqs. (\ref{R_GF})-(\ref{A_GF}), do not have a
temperature dependence.

In connection with practical applications of our results we note that in
general Majorana tunneling phases may also be used to control qubits
\cite{Gau_2020,Gau_2020a} where Majorana dark states and spaces are stabilized
and manipulated by means of special driven dissipative protocols. One may
consider those protocols as a natural subsequent stage following the Majorana
tunneling phase tuning proposed here to set up a proper initial Majorana
equilibrium state of the qubit. We would like to emphasize that in this
practical application the couplings between quantum dots and MBSs have been
fabricated purposefully to design proper functioning of the qubit via tuning
Majorana tunneling phases. As can be seen in Fig. 2 of Ref. \cite{Gau_2020},
the strengths of these couplings even have the same ratio as in our
Figs. \ref{figure_4}-\ref{figure_6}, that is $|\eta_2|/|\eta_1|=0.1$. Thus,
although at present it may be difficult to control Majorana tunneling phases
in an experiment, such a control will be necessary to manipulate Majorana
qubits in future implementations of topologically protected quantum computing
devices. Therefore it is also a quantum computing technology challenge for
experiments to tune Majorana tunneling phases via, for example, local gates
controlling the tunneling barriers between topological superconductors and
quantum dots as suggested in Refs. \cite{Gau_2020,Gau_2020a}.

Finally, as an outlook, it would be interesting, both from theoretical and
experimental points of view, to consider the case of a quantum dot with more
energy levels, two or more quantum dots coupled to a topological
superconductor or a topological superconductor which is not grounded (floating
topological superconductor). In such setups interaction between electrons
might become crucial. In particular, Coulomb blockade effects will become
important and one may explore the entropy of interacting systems in the limits
of a relatively large \cite{Fu_2010} or small \cite{Zazunov_2011,Huetzen_2012}
value of the charging energy. For quantum dots with many energy levels one
might expect that in the space of the Hamiltonian parameters there could arise
a much richer structure of the domains where the system's entropy reaches the
Majorana fractional values. The problem with two quantum dots coupled to a
topological superconductor might be useful to study the question about the
influence of a spurious tunnel coupling to any other nearby trapped state
which could impact our results. In such a setup one can use gate voltages to
change the relative position of the two energy levels. When the energy level
in one of the two quantum dots is much lower than the energy level in the
second quantum dot, one expects that the system will effectively behave as
having only one quantum dot with the lowest energy level. Thus using a gate
voltage to lower the energy level of the quantum dot one may significantly
reduce a possible impact of nearby trapped states on our results. For a
floating topological superconductor a weak Coulomb blockade leads to a smooth
change of the linear conductance. From the comparison between the entropy and
linear conductance in the lower panel of Fig. \ref{figure_5} one might also
expect for weakly Coulomb blockaded topological superconductors a smooth
change of the entropy. However, when Majorana tunneling phases and charging
energies are both involved in formation of quantum ground states, their
interplay may lead to various phenomena, like specific interaction induced
values of the tunneling phases at which the Majorana entropy is reached. In
any case, calculating the entropy is a promising approach which will uniquely
reveal the Majorana nature of thermodynamic and transport properties of
nanoscopic systems involving Majorana tunneling phases, more energy levels and
floating topological superconductors.


\begin{thebibliography}{41}%
\makeatletter
\providecommand \@ifxundefined [1]{%
 \@ifx{#1\undefined}
}%
\providecommand \@ifnum [1]{%
 \ifnum #1\expandafter \@firstoftwo
 \else \expandafter \@secondoftwo
 \fi
}%
\providecommand \@ifx [1]{%
 \ifx #1\expandafter \@firstoftwo
 \else \expandafter \@secondoftwo
 \fi
}%
\providecommand \natexlab [1]{#1}%
\providecommand \enquote  [1]{``#1''}%
\providecommand \bibnamefont  [1]{#1}%
\providecommand \bibfnamefont [1]{#1}%
\providecommand \citenamefont [1]{#1}%
\providecommand \href@noop [0]{\@secondoftwo}%
\providecommand \href [0]{\begingroup \@sanitize@url \@href}%
\providecommand \@href[1]{\@@startlink{#1}\@@href}%
\providecommand \@@href[1]{\endgroup#1\@@endlink}%
\providecommand \@sanitize@url [0]{\catcode `\\12\catcode `\$12\catcode
  `\&12\catcode `\#12\catcode `\^12\catcode `\_12\catcode `\%12\relax}%
\providecommand \@@startlink[1]{}%
\providecommand \@@endlink[0]{}%
\providecommand \url  [0]{\begingroup\@sanitize@url \@url }%
\providecommand \@url [1]{\endgroup\@href {#1}{\urlprefix }}%
\providecommand \urlprefix  [0]{URL }%
\providecommand \Eprint [0]{\href }%
\providecommand \doibase [0]{http://dx.doi.org/}%
\providecommand \selectlanguage [0]{\@gobble}%
\providecommand \bibinfo  [0]{\@secondoftwo}%
\providecommand \bibfield  [0]{\@secondoftwo}%
\providecommand \translation [1]{[#1]}%
\providecommand \BibitemOpen [0]{}%
\providecommand \bibitemStop [0]{}%
\providecommand \bibitemNoStop [0]{.\EOS\space}%
\providecommand \EOS [0]{\spacefactor3000\relax}%
\providecommand \BibitemShut  [1]{\csname bibitem#1\endcsname}%
\let\auto@bib@innerbib\@empty
\bibitem [{\citenamefont {Alicea}(2012)}]{Alicea_2012}%
  \BibitemOpen
  \bibfield  {author} {\bibinfo {author} {\bibfnamefont {J.}~\bibnamefont
  {Alicea}},\ }\bibfield  {title} {\enquote {\bibinfo {title} {New directions
  in the pursuit of {M}ajorana fermions in solid state systems},}\ }\href@noop
  {} {\bibfield  {journal} {\bibinfo  {journal} {Rep. Prog. Phys.}\ }\textbf
  {\bibinfo {volume} {75}},\ \bibinfo {pages} {076501} (\bibinfo {year}
  {2012})}\BibitemShut {NoStop}%
\bibitem [{\citenamefont {Leijnse}\ and\ \citenamefont
  {Flensberg}(2012)}]{Flensberg_2012}%
  \BibitemOpen
  \bibfield  {author} {\bibinfo {author} {\bibfnamefont {M.}~\bibnamefont
  {Leijnse}}\ and\ \bibinfo {author} {\bibfnamefont {K.}~\bibnamefont
  {Flensberg}},\ }\bibfield  {title} {\enquote {\bibinfo {title} {Introduction
  to topological superconductivity and {M}ajorana fermions},}\ }\href@noop {}
  {\bibfield  {journal} {\bibinfo  {journal} {Semicond. Sci. Technol.}\
  }\textbf {\bibinfo {volume} {27}},\ \bibinfo {pages} {124003} (\bibinfo
  {year} {2012})}\BibitemShut {NoStop}%
\bibitem [{\citenamefont {Sato}\ and\ \citenamefont
  {Fujimoto}(2016)}]{Sato_2016}%
  \BibitemOpen
  \bibfield  {author} {\bibinfo {author} {\bibfnamefont {M.}~\bibnamefont
  {Sato}}\ and\ \bibinfo {author} {\bibfnamefont {S.}~\bibnamefont
  {Fujimoto}},\ }\bibfield  {title} {\enquote {\bibinfo {title} {Majorana
  fermions and topology in superconductors},}\ }\href@noop {} {\bibfield
  {journal} {\bibinfo  {journal} {J. Phys. Soc. Japan}\ }\textbf {\bibinfo
  {volume} {85}},\ \bibinfo {pages} {072001} (\bibinfo {year}
  {2016})}\BibitemShut {NoStop}%
\bibitem [{\citenamefont {Aguado}(2017)}]{Aguado_2017}%
  \BibitemOpen
  \bibfield  {author} {\bibinfo {author} {\bibfnamefont {R.}~\bibnamefont
  {Aguado}},\ }\bibfield  {title} {\enquote {\bibinfo {title} {Majorana
  quasiparticles in condensed matter},}\ }\href@noop {} {\bibfield  {journal}
  {\bibinfo  {journal} {La Rivista del Nuovo Cimento}\ }\textbf {\bibinfo
  {volume} {40}},\ \bibinfo {pages} {523} (\bibinfo {year} {2017})}\BibitemShut
  {NoStop}%
\bibitem [{\citenamefont {Lutchyn}\ \emph {et~al.}(2018)\citenamefont
  {Lutchyn}, \citenamefont {Bakkers}, \citenamefont {Kouwenhoven},
  \citenamefont {Krogstrup}, \citenamefont {Marcus},\ and\ \citenamefont
  {Oreg}}]{Lutchyn_2018}%
  \BibitemOpen
  \bibfield  {author} {\bibinfo {author} {\bibfnamefont {R.~M.}\ \bibnamefont
  {Lutchyn}}, \bibinfo {author} {\bibfnamefont {E.~P. A.~M.}\ \bibnamefont
  {Bakkers}}, \bibinfo {author} {\bibfnamefont {L.~P.}\ \bibnamefont
  {Kouwenhoven}}, \bibinfo {author} {\bibfnamefont {P.}~\bibnamefont
  {Krogstrup}}, \bibinfo {author} {\bibfnamefont {C.~M.}\ \bibnamefont
  {Marcus}}, \ and\ \bibinfo {author} {\bibfnamefont {Y.}~\bibnamefont
  {Oreg}},\ }\bibfield  {title} {\enquote {\bibinfo {title} {Majorana zero
  modes in superconductor-semiconductor heterostructures},}\ }\href@noop {}
  {\bibfield  {journal} {\bibinfo  {journal} {Nat. Rev. Mater.}\ }\textbf
  {\bibinfo {volume} {3}},\ \bibinfo {pages} {52} (\bibinfo {year}
  {2018})}\BibitemShut {NoStop}%
\bibitem [{\citenamefont {\text{Yu.} Kitaev}(2001)}]{Kitaev_2001}%
  \BibitemOpen
  \bibfield  {author} {\bibinfo {author} {\bibfnamefont {A.}~\bibnamefont
  {\text{Yu.} Kitaev}},\ }\bibfield  {title} {\enquote {\bibinfo {title}
  {Unpaired {M}ajorana fermions in quantum wires},}\ }\href@noop {} {\bibfield
  {journal} {\bibinfo  {journal} {Phys.-Usp.}\ }\textbf {\bibinfo {volume}
  {44}},\ \bibinfo {pages} {131} (\bibinfo {year} {2001})}\BibitemShut
  {NoStop}%
\bibitem [{\citenamefont {Lutchyn}\ \emph {et~al.}(2010)\citenamefont
  {Lutchyn}, \citenamefont {Sau},\ and\ \citenamefont {\text{Das
  Sarma}}}]{Lutchyn_2010}%
  \BibitemOpen
  \bibfield  {author} {\bibinfo {author} {\bibfnamefont {R.~M.}\ \bibnamefont
  {Lutchyn}}, \bibinfo {author} {\bibfnamefont {J.~D.}\ \bibnamefont {Sau}}, \
  and\ \bibinfo {author} {\bibfnamefont {S.}~\bibnamefont {\text{Das Sarma}}},\
  }\bibfield  {title} {\enquote {\bibinfo {title} {Majorana fermions and a
  topological phase transition in semiconductor-superconductor
  heterostructures},}\ }\href@noop {} {\bibfield  {journal} {\bibinfo
  {journal} {Phys.\ Rev.\ Lett.}\ }\textbf {\bibinfo {volume} {105}},\ \bibinfo
  {pages} {077001} (\bibinfo {year} {2010})}\BibitemShut {NoStop}%
\bibitem [{\citenamefont {Oreg}\ \emph {et~al.}(2010)\citenamefont {Oreg},
  \citenamefont {Refael},\ and\ \citenamefont {von Oppen}}]{Oreg_2010}%
  \BibitemOpen
  \bibfield  {author} {\bibinfo {author} {\bibfnamefont {Y.}~\bibnamefont
  {Oreg}}, \bibinfo {author} {\bibfnamefont {G.}~\bibnamefont {Refael}}, \ and\
  \bibinfo {author} {\bibfnamefont {F.}~\bibnamefont {von Oppen}},\ }\bibfield
  {title} {\enquote {\bibinfo {title} {Helical liquids and {M}ajorana bound
  states in quantum wires},}\ }\href@noop {} {\bibfield  {journal} {\bibinfo
  {journal} {Phys.\ Rev.\ Lett.}\ }\textbf {\bibinfo {volume} {105}},\ \bibinfo
  {pages} {177002} (\bibinfo {year} {2010})}\BibitemShut {NoStop}%
\bibitem [{\citenamefont {Mourik}\ \emph {et~al.}(2012)\citenamefont {Mourik},
  \citenamefont {Zuo}, \citenamefont {Frolov}, \citenamefont {Plissard},
  \citenamefont {Bakkers},\ and\ \citenamefont {Kouwenhoven}}]{Mourik_2012}%
  \BibitemOpen
  \bibfield  {author} {\bibinfo {author} {\bibfnamefont {V.}~\bibnamefont
  {Mourik}}, \bibinfo {author} {\bibfnamefont {K.}~\bibnamefont {Zuo}},
  \bibinfo {author} {\bibfnamefont {S.~M.}\ \bibnamefont {Frolov}}, \bibinfo
  {author} {\bibfnamefont {S.~R.}\ \bibnamefont {Plissard}}, \bibinfo {author}
  {\bibfnamefont {E.~P. A.~M.}\ \bibnamefont {Bakkers}}, \ and\ \bibinfo
  {author} {\bibfnamefont {L.~P.}\ \bibnamefont {Kouwenhoven}},\ }\bibfield
  {title} {\enquote {\bibinfo {title} {Signatures of {M}ajorana fermions in
  hybrid superconductor-semiconductor nanowire devices},}\ }\href@noop {}
  {\bibfield  {journal} {\bibinfo  {journal} {Science}\ }\textbf {\bibinfo
  {volume} {336}},\ \bibinfo {pages} {1003} (\bibinfo {year}
  {2012})}\BibitemShut {NoStop}%
\bibitem [{\citenamefont {Albrecht}\ \emph {et~al.}(2016)\citenamefont
  {Albrecht}, \citenamefont {Higginbotham}, \citenamefont {Madsen},
  \citenamefont {Kuemmeth}, \citenamefont {Jespersen}, \citenamefont
  {Nyg{\r{a}}rd}, \citenamefont {Krogstrup},\ and\ \citenamefont
  {Marcus}}]{Albrecht_2016}%
  \BibitemOpen
  \bibfield  {author} {\bibinfo {author} {\bibfnamefont {S.~M.}\ \bibnamefont
  {Albrecht}}, \bibinfo {author} {\bibfnamefont {A.~P.}\ \bibnamefont
  {Higginbotham}}, \bibinfo {author} {\bibfnamefont {M.}~\bibnamefont
  {Madsen}}, \bibinfo {author} {\bibfnamefont {F.}~\bibnamefont {Kuemmeth}},
  \bibinfo {author} {\bibfnamefont {T.~S.}\ \bibnamefont {Jespersen}}, \bibinfo
  {author} {\bibfnamefont {J.}~\bibnamefont {Nyg{\r{a}}rd}}, \bibinfo {author}
  {\bibfnamefont {P.}~\bibnamefont {Krogstrup}}, \ and\ \bibinfo {author}
  {\bibfnamefont {C.~M.}\ \bibnamefont {Marcus}},\ }\bibfield  {title}
  {\enquote {\bibinfo {title} {Exponential protection of zero modes in
  {M}ajorana islands},}\ }\href@noop {} {\bibfield  {journal} {\bibinfo
  {journal} {Nature}\ }\textbf {\bibinfo {volume} {531}},\ \bibinfo {pages}
  {206} (\bibinfo {year} {2016})}\BibitemShut {NoStop}%
\bibitem [{\citenamefont {Zhang}\ \emph {et~al.}(2018)\citenamefont {Zhang},
  \citenamefont {Liu}, \citenamefont {Gazibegovic}, \citenamefont {Xu},
  \citenamefont {Logan}, \citenamefont {Wang}, \citenamefont {van Loo},
  \citenamefont {Bommer}, \citenamefont {de~Moor}, \citenamefont {Car},
  \citenamefont {het Veld}, \citenamefont {van Veldhoven}, \citenamefont
  {Koelling}, \citenamefont {Verheijen}, \citenamefont {Pendharkar},
  \citenamefont {Pennachio}, \citenamefont {Shojaei}, \citenamefont {Lee},
  \citenamefont {Palmstr{\o}m}, \citenamefont {Bakkers}, \citenamefont
  {Sarma},\ and\ \citenamefont {Kouwenhoven}}]{Zhang_2018}%
  \BibitemOpen
  \bibfield  {author} {\bibinfo {author} {\bibfnamefont {H.}~\bibnamefont
  {Zhang}}, \bibinfo {author} {\bibfnamefont {C.-X.}\ \bibnamefont {Liu}},
  \bibinfo {author} {\bibfnamefont {S.}~\bibnamefont {Gazibegovic}}, \bibinfo
  {author} {\bibfnamefont {D.}~\bibnamefont {Xu}}, \bibinfo {author}
  {\bibfnamefont {J.~A.}\ \bibnamefont {Logan}}, \bibinfo {author}
  {\bibfnamefont {G.}~\bibnamefont {Wang}}, \bibinfo {author} {\bibfnamefont
  {N.}~\bibnamefont {van Loo}}, \bibinfo {author} {\bibfnamefont {J.~D.~S.}\
  \bibnamefont {Bommer}}, \bibinfo {author} {\bibfnamefont {M.~W.~A.}\
  \bibnamefont {de~Moor}}, \bibinfo {author} {\bibfnamefont {D.}~\bibnamefont
  {Car}}, \bibinfo {author} {\bibfnamefont {R.~L. M.~O.}\ \bibnamefont {het
  Veld}}, \bibinfo {author} {\bibfnamefont {P.~J.}\ \bibnamefont {van
  Veldhoven}}, \bibinfo {author} {\bibfnamefont {S.}~\bibnamefont {Koelling}},
  \bibinfo {author} {\bibfnamefont {M.~A.}\ \bibnamefont {Verheijen}}, \bibinfo
  {author} {\bibfnamefont {M.}~\bibnamefont {Pendharkar}}, \bibinfo {author}
  {\bibfnamefont {D.~J.}\ \bibnamefont {Pennachio}}, \bibinfo {author}
  {\bibfnamefont {B.}~\bibnamefont {Shojaei}}, \bibinfo {author} {\bibfnamefont
  {J.~S.}\ \bibnamefont {Lee}}, \bibinfo {author} {\bibfnamefont {C.~J.}\
  \bibnamefont {Palmstr{\o}m}}, \bibinfo {author} {\bibfnamefont {E.~P. A.~M.}\
  \bibnamefont {Bakkers}}, \bibinfo {author} {\bibfnamefont {S.~D.}\
  \bibnamefont {Sarma}}, \ and\ \bibinfo {author} {\bibfnamefont {L.~P.}\
  \bibnamefont {Kouwenhoven}},\ }\bibfield  {title} {\enquote {\bibinfo {title}
  {Quantized {M}ajorana conductance},}\ }\href@noop {} {\bibfield  {journal}
  {\bibinfo  {journal} {Nature}\ }\textbf {\bibinfo {volume} {556}},\ \bibinfo
  {pages} {74} (\bibinfo {year} {2018})}\BibitemShut {NoStop}%
\bibitem [{\citenamefont {Leijnse}(2014)}]{Leijnse_2014}%
  \BibitemOpen
  \bibfield  {author} {\bibinfo {author} {\bibfnamefont {M.}~\bibnamefont
  {Leijnse}},\ }\bibfield  {title} {\enquote {\bibinfo {title} {Thermoelectric
  signatures of a {M}ajorana bound state coupled to a quantum dot},}\
  }\href@noop {} {\bibfield  {journal} {\bibinfo  {journal} {New J. Phys.}\
  }\textbf {\bibinfo {volume} {16}},\ \bibinfo {pages} {015029} (\bibinfo
  {year} {2014})}\BibitemShut {NoStop}%
\bibitem [{\citenamefont {Ramos-Andrade}\ \emph {et~al.}(2016)\citenamefont
  {Ramos-Andrade}, \citenamefont {{\'Avalos-Ovando}}, \citenamefont
  {Orellana},\ and\ \citenamefont {Ulloa}}]{Ramos-Andrade_2016}%
  \BibitemOpen
  \bibfield  {author} {\bibinfo {author} {\bibfnamefont {J.~P.}\ \bibnamefont
  {Ramos-Andrade}}, \bibinfo {author} {\bibfnamefont {O.}~\bibnamefont
  {{\'Avalos-Ovando}}}, \bibinfo {author} {\bibfnamefont {P.~A.}\ \bibnamefont
  {Orellana}}, \ and\ \bibinfo {author} {\bibfnamefont {S.~E.}\ \bibnamefont
  {Ulloa}},\ }\bibfield  {title} {\enquote {\bibinfo {title} {Thermoelectric
  transport through {M}ajorana bound states and violation of
  {W}iedemann-{F}ranz law},}\ }\href@noop {} {\bibfield  {journal} {\bibinfo
  {journal} {Phys.\ Rev.\ B}\ }\textbf {\bibinfo {volume} {94}},\ \bibinfo
  {pages} {155436} (\bibinfo {year} {2016})}\BibitemShut {NoStop}%
\bibitem [{\citenamefont {Hong}\ \emph {et~al.}(2020)\citenamefont {Hong},
  \citenamefont {Chi}, \citenamefont {Fu}, \citenamefont {Hou}, \citenamefont
  {Wang}, \citenamefont {Li}, \citenamefont {Liu}, \citenamefont {Yao},\ and\
  \citenamefont {Zhang}}]{Hong_2020}%
  \BibitemOpen
  \bibfield  {author} {\bibinfo {author} {\bibfnamefont {L.}~\bibnamefont
  {Hong}}, \bibinfo {author} {\bibfnamefont {F.}~\bibnamefont {Chi}}, \bibinfo
  {author} {\bibfnamefont {Z.-G.}\ \bibnamefont {Fu}}, \bibinfo {author}
  {\bibfnamefont {Y.-F.}\ \bibnamefont {Hou}}, \bibinfo {author} {\bibfnamefont
  {Z.}~\bibnamefont {Wang}}, \bibinfo {author} {\bibfnamefont {K.-M.}\
  \bibnamefont {Li}}, \bibinfo {author} {\bibfnamefont {J.}~\bibnamefont
  {Liu}}, \bibinfo {author} {\bibfnamefont {H.}~\bibnamefont {Yao}}, \ and\
  \bibinfo {author} {\bibfnamefont {P.}~\bibnamefont {Zhang}},\ }\bibfield
  {title} {\enquote {\bibinfo {title} {Large enhancement of thermoelectric
  effect by {M}ajorana bound states coupled to a quantum dot},}\ }\href@noop {}
  {\bibfield  {journal} {\bibinfo  {journal} {J. Appl. Phys.}\ }\textbf
  {\bibinfo {volume} {127}},\ \bibinfo {pages} {124302} (\bibinfo {year}
  {2020})}\BibitemShut {NoStop}%
\bibitem [{\citenamefont {Chi}\ \emph {et~al.}(2020{\natexlab{a}})\citenamefont
  {Chi}, \citenamefont {Fu}, \citenamefont {Liu}, \citenamefont {Li},
  \citenamefont {Wang},\ and\ \citenamefont {Zhang}}]{Chi_2020}%
  \BibitemOpen
  \bibfield  {author} {\bibinfo {author} {\bibfnamefont {F.}~\bibnamefont
  {Chi}}, \bibinfo {author} {\bibfnamefont {Z.-G.}\ \bibnamefont {Fu}},
  \bibinfo {author} {\bibfnamefont {J.}~\bibnamefont {Liu}}, \bibinfo {author}
  {\bibfnamefont {K.-M.}\ \bibnamefont {Li}}, \bibinfo {author} {\bibfnamefont
  {Z.}~\bibnamefont {Wang}}, \ and\ \bibinfo {author} {\bibfnamefont
  {P.}~\bibnamefont {Zhang}},\ }\bibfield  {title} {\enquote {\bibinfo {title}
  {Thermoelectric effect in a correlated quantum dot side-coupled to {M}ajorana
  bound states},}\ }\href@noop {} {\bibfield  {journal} {\bibinfo  {journal}
  {Nanoscale Res. Lett.}\ }\textbf {\bibinfo {volume} {15}},\ \bibinfo {pages}
  {79} (\bibinfo {year} {2020}{\natexlab{a}})}\BibitemShut {NoStop}%
\bibitem [{\citenamefont {Bondyopadhaya}\ and\ \citenamefont
  {Roy}(2020)}]{Bondyopadhaya_2020}%
  \BibitemOpen
  \bibfield  {author} {\bibinfo {author} {\bibfnamefont {N.}~\bibnamefont
  {Bondyopadhaya}}\ and\ \bibinfo {author} {\bibfnamefont {D.}~\bibnamefont
  {Roy}},\ }\bibfield  {title} {\enquote {\bibinfo {title} {Nonequilibrium
  electrical, thermal and spin transport in open quantum systems of topological
  superconductors, semiconductors and metals},}\ }\href@noop {} {\bibfield
  {journal} {\bibinfo  {journal} {arXiv:2010.08336v1}\ } (\bibinfo {year}
  {2020})}\BibitemShut {NoStop}%
\bibitem [{\citenamefont {Liu}\ \emph {et~al.}(2015{\natexlab{a}})\citenamefont
  {Liu}, \citenamefont {Cheng},\ and\ \citenamefont {Lutchyn}}]{Liu_2015}%
  \BibitemOpen
  \bibfield  {author} {\bibinfo {author} {\bibfnamefont {D.~E.}\ \bibnamefont
  {Liu}}, \bibinfo {author} {\bibfnamefont {M.}~\bibnamefont {Cheng}}, \ and\
  \bibinfo {author} {\bibfnamefont {R.~M.}\ \bibnamefont {Lutchyn}},\
  }\bibfield  {title} {\enquote {\bibinfo {title} {Probing {M}ajorana physics
  in quantum-dot shot-noise experiments},}\ }\href@noop {} {\bibfield
  {journal} {\bibinfo  {journal} {Phys.\ Rev.\ B}\ }\textbf {\bibinfo {volume}
  {91}},\ \bibinfo {pages} {081405(R)} (\bibinfo {year}
  {2015}{\natexlab{a}})}\BibitemShut {NoStop}%
\bibitem [{\citenamefont {Liu}\ \emph {et~al.}(2015{\natexlab{b}})\citenamefont
  {Liu}, \citenamefont {Levchenko},\ and\ \citenamefont {Lutchyn}}]{Liu_2015a}%
  \BibitemOpen
  \bibfield  {author} {\bibinfo {author} {\bibfnamefont {D.~E.}\ \bibnamefont
  {Liu}}, \bibinfo {author} {\bibfnamefont {A.}~\bibnamefont {Levchenko}}, \
  and\ \bibinfo {author} {\bibfnamefont {R.~M.}\ \bibnamefont {Lutchyn}},\
  }\bibfield  {title} {\enquote {\bibinfo {title} {Majorana zero modes choose
  {E}uler numbers as revealed by full counting statistics},}\ }\href@noop {}
  {\bibfield  {journal} {\bibinfo  {journal} {Phys.\ Rev.\ B}\ }\textbf
  {\bibinfo {volume} {92}},\ \bibinfo {pages} {205422} (\bibinfo {year}
  {2015}{\natexlab{b}})}\BibitemShut {NoStop}%
\bibitem [{\citenamefont {Haim}\ \emph {et~al.}(2015)\citenamefont {Haim},
  \citenamefont {Berg}, \citenamefont {von Oppen},\ and\ \citenamefont
  {Oreg}}]{Haim_2015}%
  \BibitemOpen
  \bibfield  {author} {\bibinfo {author} {\bibfnamefont {A.}~\bibnamefont
  {Haim}}, \bibinfo {author} {\bibfnamefont {E.}~\bibnamefont {Berg}}, \bibinfo
  {author} {\bibfnamefont {F.}~\bibnamefont {von Oppen}}, \ and\ \bibinfo
  {author} {\bibfnamefont {Y.}~\bibnamefont {Oreg}},\ }\bibfield  {title}
  {\enquote {\bibinfo {title} {Current correlations in a {M}ajorana beam
  splitter},}\ }\href@noop {} {\bibfield  {journal} {\bibinfo  {journal}
  {Phys.\ Rev.\ B}\ }\textbf {\bibinfo {volume} {92}},\ \bibinfo {pages}
  {245112} (\bibinfo {year} {2015})}\BibitemShut {NoStop}%
\bibitem [{\citenamefont {Valentini}\ \emph {et~al.}(2016)\citenamefont
  {Valentini}, \citenamefont {Governale}, \citenamefont {Fazio},\ and\
  \citenamefont {Taddei}}]{Valentini_2016}%
  \BibitemOpen
  \bibfield  {author} {\bibinfo {author} {\bibfnamefont {S.}~\bibnamefont
  {Valentini}}, \bibinfo {author} {\bibfnamefont {M.}~\bibnamefont
  {Governale}}, \bibinfo {author} {\bibfnamefont {R.}~\bibnamefont {Fazio}}, \
  and\ \bibinfo {author} {\bibfnamefont {F.}~\bibnamefont {Taddei}},\
  }\bibfield  {title} {\enquote {\bibinfo {title} {Finite-frequency noise in a
  topological superconducting wire},}\ }\href@noop {} {\bibfield  {journal}
  {\bibinfo  {journal} {Physica E}\ }\textbf {\bibinfo {volume} {75}},\
  \bibinfo {pages} {15} (\bibinfo {year} {2016})}\BibitemShut {NoStop}%
\bibitem [{\citenamefont {Smirnov}(2017)}]{Smirnov_2017}%
  \BibitemOpen
  \bibfield  {author} {\bibinfo {author} {\bibfnamefont {S.}~\bibnamefont
  {Smirnov}},\ }\bibfield  {title} {\enquote {\bibinfo {title} {Non-equilibrium
  {M}ajorana fluctuations},}\ }\href@noop {} {\bibfield  {journal} {\bibinfo
  {journal} {New J. Phys.}\ }\textbf {\bibinfo {volume} {19}},\ \bibinfo
  {pages} {063020} (\bibinfo {year} {2017})}\BibitemShut {NoStop}%
\bibitem [{\citenamefont {Bathellier}\ \emph {et~al.}(2019)\citenamefont
  {Bathellier}, \citenamefont {Raymond}, \citenamefont {Jonckheere},
  \citenamefont {Rech}, \citenamefont {Zazunov},\ and\ \citenamefont
  {Martin}}]{Bathellier_2019}%
  \BibitemOpen
  \bibfield  {author} {\bibinfo {author} {\bibfnamefont {D.}~\bibnamefont
  {Bathellier}}, \bibinfo {author} {\bibfnamefont {L.}~\bibnamefont {Raymond}},
  \bibinfo {author} {\bibfnamefont {T.}~\bibnamefont {Jonckheere}}, \bibinfo
  {author} {\bibfnamefont {J.}~\bibnamefont {Rech}}, \bibinfo {author}
  {\bibfnamefont {A.}~\bibnamefont {Zazunov}}, \ and\ \bibinfo {author}
  {\bibfnamefont {T.}~\bibnamefont {Martin}},\ }\bibfield  {title} {\enquote
  {\bibinfo {title} {Finite frequency noise in a normal metal - topological
  superconductor junction},}\ }\href@noop {} {\bibfield  {journal} {\bibinfo
  {journal} {Phys.\ Rev.\ B}\ }\textbf {\bibinfo {volume} {99}},\ \bibinfo
  {pages} {104502} (\bibinfo {year} {2019})}\BibitemShut {NoStop}%
\bibitem [{\citenamefont {Smirnov}(2019{\natexlab{a}})}]{Smirnov_2019}%
  \BibitemOpen
  \bibfield  {author} {\bibinfo {author} {\bibfnamefont {S.}~\bibnamefont
  {Smirnov}},\ }\bibfield  {title} {\enquote {\bibinfo {title} {Majorana
  finite-frequency nonequilibrium quantum noise},}\ }\href@noop {} {\bibfield
  {journal} {\bibinfo  {journal} {Phys.\ Rev.\ B}\ }\textbf {\bibinfo {volume}
  {99}},\ \bibinfo {pages} {165427} (\bibinfo {year}
  {2019}{\natexlab{a}})}\BibitemShut {NoStop}%
\bibitem [{\citenamefont {Smirnov}(2018)}]{Smirnov_2018}%
  \BibitemOpen
  \bibfield  {author} {\bibinfo {author} {\bibfnamefont {S.}~\bibnamefont
  {Smirnov}},\ }\bibfield  {title} {\enquote {\bibinfo {title} {Universal
  {M}ajorana thermoelectric noise},}\ }\href@noop {} {\bibfield  {journal}
  {\bibinfo  {journal} {Phys.\ Rev.\ B}\ }\textbf {\bibinfo {volume} {97}},\
  \bibinfo {pages} {165434} (\bibinfo {year} {2018})}\BibitemShut {NoStop}%
\bibitem [{\citenamefont {Smirnov}(2019{\natexlab{b}})}]{Smirnov_2019a}%
  \BibitemOpen
  \bibfield  {author} {\bibinfo {author} {\bibfnamefont {S.}~\bibnamefont
  {Smirnov}},\ }\bibfield  {title} {\enquote {\bibinfo {title} {Dynamic
  {M}ajorana resonances and universal symmetry of nonequilibrium thermoelectric
  quantum noise},}\ }\href@noop {} {\bibfield  {journal} {\bibinfo  {journal}
  {Phys.\ Rev.\ B}\ }\textbf {\bibinfo {volume} {100}},\ \bibinfo {pages}
  {245410} (\bibinfo {year} {2019}{\natexlab{b}})}\BibitemShut {NoStop}%
\bibitem [{\citenamefont {Tang}\ and\ \citenamefont {Mao}(2020)}]{Tang_2020}%
  \BibitemOpen
  \bibfield  {author} {\bibinfo {author} {\bibfnamefont {L.-W.}\ \bibnamefont
  {Tang}}\ and\ \bibinfo {author} {\bibfnamefont {W.-G.}\ \bibnamefont {Mao}},\
  }\bibfield  {title} {\enquote {\bibinfo {title} {Detection of {M}ajorana
  bound states by sign change of the tunnel magnetoresistance in a quantum dot
  coupled to ferromagnetic electrodes},}\ }\href@noop {} {\bibfield  {journal}
  {\bibinfo  {journal} {Front. Phys.}\ }\textbf {\bibinfo {volume} {8}},\
  \bibinfo {pages} {147} (\bibinfo {year} {2020})}\BibitemShut {NoStop}%
\bibitem [{\citenamefont {Zhang}\ and\ \citenamefont
  {Sp\r{a}nsl{\"a}tt}(2020)}]{Zhang_2020}%
  \BibitemOpen
  \bibfield  {author} {\bibinfo {author} {\bibfnamefont {G.}~\bibnamefont
  {Zhang}}\ and\ \bibinfo {author} {\bibfnamefont {C.}~\bibnamefont
  {Sp\r{a}nsl{\"a}tt}},\ }\bibfield  {title} {\enquote {\bibinfo {title}
  {Distinguishing between topological and quasi {M}ajorana zero modes with a
  dissipative resonant level},}\ }\href@noop {} {\bibfield  {journal} {\bibinfo
   {journal} {Phys.\ Rev.\ B}\ }\textbf {\bibinfo {volume} {102}},\ \bibinfo
  {pages} {045111} (\bibinfo {year} {2020})}\BibitemShut {NoStop}%
\bibitem [{\citenamefont {Chi}\ \emph {et~al.}(2020{\natexlab{b}})\citenamefont
  {Chi}, \citenamefont {He}, \citenamefont {Wang}, \citenamefont {Fu},
  \citenamefont {Liu}, \citenamefont {Liu},\ and\ \citenamefont
  {Zhang}}]{Chi_2020a}%
  \BibitemOpen
  \bibfield  {author} {\bibinfo {author} {\bibfnamefont {F.}~\bibnamefont
  {Chi}}, \bibinfo {author} {\bibfnamefont {T.-Y.}\ \bibnamefont {He}},
  \bibinfo {author} {\bibfnamefont {J.}~\bibnamefont {Wang}}, \bibinfo {author}
  {\bibfnamefont {Z.-G.}\ \bibnamefont {Fu}}, \bibinfo {author} {\bibfnamefont
  {L.-M.}\ \bibnamefont {Liu}}, \bibinfo {author} {\bibfnamefont
  {P.}~\bibnamefont {Liu}}, \ and\ \bibinfo {author} {\bibfnamefont
  {P.}~\bibnamefont {Zhang}},\ }\bibfield  {title} {\enquote {\bibinfo {title}
  {Photon-assisted transport through a quantum dot side-coupled to {M}ajorana
  bound states},}\ }\href@noop {} {\bibfield  {journal} {\bibinfo  {journal}
  {Front. Phys.}\ }\textbf {\bibinfo {volume} {8}},\ \bibinfo {pages} {254}
  (\bibinfo {year} {2020}{\natexlab{b}})}\BibitemShut {NoStop}%
\bibitem [{\citenamefont {Simons}\ \emph {et~al.}(2020)\citenamefont {Simons},
  \citenamefont {Meidan},\ and\ \citenamefont {Romito}}]{Simons_2020}%
  \BibitemOpen
  \bibfield  {author} {\bibinfo {author} {\bibfnamefont {T.}~\bibnamefont
  {Simons}}, \bibinfo {author} {\bibfnamefont {D.}~\bibnamefont {Meidan}}, \
  and\ \bibinfo {author} {\bibfnamefont {A.}~\bibnamefont {Romito}},\
  }\bibfield  {title} {\enquote {\bibinfo {title} {Pumped heat and charge
  statistics from {M}ajorana braiding},}\ }\href@noop {} {\bibfield  {journal}
  {\bibinfo  {journal} {arXiv:2005.11727v1}\ } (\bibinfo {year}
  {2020})}\BibitemShut {NoStop}%
\bibitem [{\citenamefont {Smirnov}(2020)}]{Smirnov_2020a}%
  \BibitemOpen
  \bibfield  {author} {\bibinfo {author} {\bibfnamefont {S.}~\bibnamefont
  {Smirnov}},\ }\bibfield  {title} {\enquote {\bibinfo {title} {Dual {M}ajorana
  universality in thermally induced nonequilibrium},}\ }\href@noop {}
  {\bibfield  {journal} {\bibinfo  {journal} {Phys.\ Rev.\ B}\ }\textbf
  {\bibinfo {volume} {101}},\ \bibinfo {pages} {125417} (\bibinfo {year}
  {2020})}\BibitemShut {NoStop}%
\bibitem [{\citenamefont {Manousakis}\ \emph {et~al.}(2020)\citenamefont
  {Manousakis}, \citenamefont {Wille}, \citenamefont {Altland}, \citenamefont
  {Egger}, \citenamefont {Flensberg},\ and\ \citenamefont
  {Hassler}}]{Manousakis_2020}%
  \BibitemOpen
  \bibfield  {author} {\bibinfo {author} {\bibfnamefont {J.}~\bibnamefont
  {Manousakis}}, \bibinfo {author} {\bibfnamefont {C.}~\bibnamefont {Wille}},
  \bibinfo {author} {\bibfnamefont {A.}~\bibnamefont {Altland}}, \bibinfo
  {author} {\bibfnamefont {R.}~\bibnamefont {Egger}}, \bibinfo {author}
  {\bibfnamefont {K.}~\bibnamefont {Flensberg}}, \ and\ \bibinfo {author}
  {\bibfnamefont {F.}~\bibnamefont {Hassler}},\ }\bibfield  {title} {\enquote
  {\bibinfo {title} {Weak measurement protocols for {M}ajorana bound state
  identification},}\ }\href@noop {} {\bibfield  {journal} {\bibinfo  {journal}
  {Phys.\ Rev.\ Lett.}\ }\textbf {\bibinfo {volume} {124}},\ \bibinfo {pages}
  {096801} (\bibinfo {year} {2020})}\BibitemShut {NoStop}%
\bibitem [{\citenamefont {Hartman}\ \emph {et~al.}(2018)\citenamefont
  {Hartman}, \citenamefont {Olsen}, \citenamefont {L{\"u}scher}, \citenamefont
  {Samani}, \citenamefont {Fallahi}, \citenamefont {Gardner}, \citenamefont
  {Manfra},\ and\ \citenamefont {Folk}}]{Hartman_2018}%
  \BibitemOpen
  \bibfield  {author} {\bibinfo {author} {\bibfnamefont {N.}~\bibnamefont
  {Hartman}}, \bibinfo {author} {\bibfnamefont {C.}~\bibnamefont {Olsen}},
  \bibinfo {author} {\bibfnamefont {S.}~\bibnamefont {L{\"u}scher}}, \bibinfo
  {author} {\bibfnamefont {M.}~\bibnamefont {Samani}}, \bibinfo {author}
  {\bibfnamefont {S.}~\bibnamefont {Fallahi}}, \bibinfo {author} {\bibfnamefont
  {G.~C.}\ \bibnamefont {Gardner}}, \bibinfo {author} {\bibfnamefont
  {M.}~\bibnamefont {Manfra}}, \ and\ \bibinfo {author} {\bibfnamefont
  {J.}~\bibnamefont {Folk}},\ }\bibfield  {title} {\enquote {\bibinfo {title}
  {Direct entropy measurement in a mesoscopic quantum system},}\ }\href@noop {}
  {\bibfield  {journal} {\bibinfo  {journal} {Nature Physics}\ }\textbf
  {\bibinfo {volume} {14}},\ \bibinfo {pages} {1083} (\bibinfo {year}
  {2018})}\BibitemShut {NoStop}%
\bibitem [{\citenamefont {Kleeorin}\ \emph {et~al.}(2019)\citenamefont
  {Kleeorin}, \citenamefont {Thierschmann}, \citenamefont {Buhmann},
  \citenamefont {Georges}, \citenamefont {Molenkamp},\ and\ \citenamefont
  {Meir}}]{Kleeorin_2019}%
  \BibitemOpen
  \bibfield  {author} {\bibinfo {author} {\bibfnamefont {Y.}~\bibnamefont
  {Kleeorin}}, \bibinfo {author} {\bibfnamefont {H.}~\bibnamefont
  {Thierschmann}}, \bibinfo {author} {\bibfnamefont {H.}~\bibnamefont
  {Buhmann}}, \bibinfo {author} {\bibfnamefont {A.}~\bibnamefont {Georges}},
  \bibinfo {author} {\bibfnamefont {L.~W.}\ \bibnamefont {Molenkamp}}, \ and\
  \bibinfo {author} {\bibfnamefont {Y.}~\bibnamefont {Meir}},\ }\bibfield
  {title} {\enquote {\bibinfo {title} {How to measure the entropy of a
  mesoscopic system via thermoelectric transport},}\ }\href@noop {} {\bibfield
  {journal} {\bibinfo  {journal} {Nat. Commun.}\ }\textbf {\bibinfo {volume}
  {10}},\ \bibinfo {pages} {5801} (\bibinfo {year} {2019})}\BibitemShut
  {NoStop}%
\bibitem [{\citenamefont {Smirnov}(2015)}]{Smirnov_2015}%
  \BibitemOpen
  \bibfield  {author} {\bibinfo {author} {\bibfnamefont {S.}~\bibnamefont
  {Smirnov}},\ }\bibfield  {title} {\enquote {\bibinfo {title} {Majorana
  tunneling entropy},}\ }\href@noop {} {\bibfield  {journal} {\bibinfo
  {journal} {Phys.\ Rev.\ B}\ }\textbf {\bibinfo {volume} {92}},\ \bibinfo
  {pages} {195312} (\bibinfo {year} {2015})}\BibitemShut {NoStop}%
\bibitem [{\citenamefont {Sela}\ \emph {et~al.}(2019)\citenamefont {Sela},
  \citenamefont {Oreg}, \citenamefont {Plugge}, \citenamefont {Hartman},
  \citenamefont {L{\"u}scher},\ and\ \citenamefont {Folk}}]{Sela_2019}%
  \BibitemOpen
  \bibfield  {author} {\bibinfo {author} {\bibfnamefont {E.}~\bibnamefont
  {Sela}}, \bibinfo {author} {\bibfnamefont {Y.}~\bibnamefont {Oreg}}, \bibinfo
  {author} {\bibfnamefont {S.}~\bibnamefont {Plugge}}, \bibinfo {author}
  {\bibfnamefont {N.}~\bibnamefont {Hartman}}, \bibinfo {author} {\bibfnamefont
  {S.}~\bibnamefont {L{\"u}scher}}, \ and\ \bibinfo {author} {\bibfnamefont
  {J.}~\bibnamefont {Folk}},\ }\bibfield  {title} {\enquote {\bibinfo {title}
  {Detecting the universal fractional entropy of {M}ajorana zero modes},}\
  }\href@noop {} {\bibfield  {journal} {\bibinfo  {journal} {Phys.\ Rev.\
  Lett.}\ }\textbf {\bibinfo {volume} {123}},\ \bibinfo {pages} {147702}
  (\bibinfo {year} {2019})}\BibitemShut {NoStop}%
\bibitem [{\citenamefont {Altland}\ and\ \citenamefont
  {Simons}(2010)}]{Altland_2010}%
  \BibitemOpen
  \bibfield  {author} {\bibinfo {author} {\bibfnamefont {A.}~\bibnamefont
  {Altland}}\ and\ \bibinfo {author} {\bibfnamefont {B.}~\bibnamefont
  {Simons}},\ }\href@noop {} {\emph {\bibinfo {title} {Condensed Matter Field
  Theory}}},\ \bibinfo {edition} {2nd}\ ed.\ (\bibinfo  {publisher} {Cambridge
  University Press, Cambridge},\ \bibinfo {year} {2010})\BibitemShut {NoStop}%
\bibitem [{\citenamefont {Gau}\ \emph {et~al.}(2020{\natexlab{a}})\citenamefont
  {Gau}, \citenamefont {Egger}, \citenamefont {Zazunov},\ and\ \citenamefont
  {Gefen}}]{Gau_2020}%
  \BibitemOpen
  \bibfield  {author} {\bibinfo {author} {\bibfnamefont {M.}~\bibnamefont
  {Gau}}, \bibinfo {author} {\bibfnamefont {R.}~\bibnamefont {Egger}}, \bibinfo
  {author} {\bibfnamefont {A.}~\bibnamefont {Zazunov}}, \ and\ \bibinfo
  {author} {\bibfnamefont {Y.}~\bibnamefont {Gefen}},\ }\bibfield  {title}
  {\enquote {\bibinfo {title} {Towards dark space stabilization and
  manipulation in driven dissipative {M}ajorana platforms},}\ }\href@noop {}
  {\bibfield  {journal} {\bibinfo  {journal} {Phys.\ Rev.\ B}\ }\textbf
  {\bibinfo {volume} {102}},\ \bibinfo {pages} {134501} (\bibinfo {year}
  {2020}{\natexlab{a}})}\BibitemShut {NoStop}%
\bibitem [{\citenamefont {Gau}\ \emph {et~al.}(2020{\natexlab{b}})\citenamefont
  {Gau}, \citenamefont {Egger}, \citenamefont {Zazunov},\ and\ \citenamefont
  {Gefen}}]{Gau_2020a}%
  \BibitemOpen
  \bibfield  {author} {\bibinfo {author} {\bibfnamefont {M.}~\bibnamefont
  {Gau}}, \bibinfo {author} {\bibfnamefont {R.}~\bibnamefont {Egger}}, \bibinfo
  {author} {\bibfnamefont {A.}~\bibnamefont {Zazunov}}, \ and\ \bibinfo
  {author} {\bibfnamefont {Yuval}\ \bibnamefont {Gefen}},\ }\bibfield  {title}
  {\enquote {\bibinfo {title} {Driven dissipative {M}ajorana dark spaces},}\
  }\href@noop {} {\bibfield  {journal} {\bibinfo  {journal} {Phys.\ Rev.\
  Lett.}\ }\textbf {\bibinfo {volume} {125}},\ \bibinfo {pages} {147701}
  (\bibinfo {year} {2020}{\natexlab{b}})}\BibitemShut {NoStop}%
\bibitem [{\citenamefont {Fu}(2010)}]{Fu_2010}%
  \BibitemOpen
  \bibfield  {author} {\bibinfo {author} {\bibfnamefont {L.}~\bibnamefont
  {Fu}},\ }\bibfield  {title} {\enquote {\bibinfo {title} {Electron
  teleportation via {M}ajorana bound states in a mesoscopic superconductor},}\
  }\href@noop {} {\bibfield  {journal} {\bibinfo  {journal} {Phys.\ Rev.\
  Lett.}\ }\textbf {\bibinfo {volume} {104}},\ \bibinfo {pages} {056402}
  (\bibinfo {year} {2010})}\BibitemShut {NoStop}%
\bibitem [{\citenamefont {Zazunov}\ \emph {et~al.}(2011)\citenamefont
  {Zazunov}, \citenamefont {Yeyati},\ and\ \citenamefont
  {Egger}}]{Zazunov_2011}%
  \BibitemOpen
  \bibfield  {author} {\bibinfo {author} {\bibfnamefont {A.}~\bibnamefont
  {Zazunov}}, \bibinfo {author} {\bibfnamefont {A.~{Levy}}\ \bibnamefont
  {Yeyati}}, \ and\ \bibinfo {author} {\bibfnamefont {R.}~\bibnamefont
  {Egger}},\ }\bibfield  {title} {\enquote {\bibinfo {title} {Coulomb blockade
  of {M}ajorana-fermion-induced transport},}\ }\href@noop {} {\bibfield
  {journal} {\bibinfo  {journal} {Phys.\ Rev.\ B}\ }\textbf {\bibinfo {volume}
  {84}},\ \bibinfo {pages} {165440} (\bibinfo {year} {2011})}\BibitemShut
  {NoStop}%
\bibitem [{\citenamefont {H{\"u}tzen}\ \emph {et~al.}(2012)\citenamefont
  {H{\"u}tzen}, \citenamefont {Zazunov}, \citenamefont {Braunecker},
  \citenamefont {Yeyati},\ and\ \citenamefont {Egger}}]{Huetzen_2012}%
  \BibitemOpen
  \bibfield  {author} {\bibinfo {author} {\bibfnamefont {R.}~\bibnamefont
  {H{\"u}tzen}}, \bibinfo {author} {\bibfnamefont {A.}~\bibnamefont {Zazunov}},
  \bibinfo {author} {\bibfnamefont {B.}~\bibnamefont {Braunecker}}, \bibinfo
  {author} {\bibfnamefont {A.~{Levy}}\ \bibnamefont {Yeyati}}, \ and\ \bibinfo
  {author} {\bibfnamefont {R.}~\bibnamefont {Egger}},\ }\bibfield  {title}
  {\enquote {\bibinfo {title} {{M}ajorana single-charge transistor},}\
  }\href@noop {} {\bibfield  {journal} {\bibinfo  {journal} {Phys.\ Rev.\
  Lett.}\ }\textbf {\bibinfo {volume} {109}},\ \bibinfo {pages} {166403}
  (\bibinfo {year} {2012})}\BibitemShut {NoStop}%
\end{thebibliography}
\end{document}